\documentclass[aip,jcp,twocolumn,10pt,amsmath,amssymb,a4paper]{revtex4-1}

\usepackage{graphicx}
\usepackage{dcolumn}
\usepackage{bm}
\usepackage{epstopdf}
\DeclareTextSymbol{\degre}{OT1}{23}

\begin{document}





\title{Theory and Monte Carlo simulations for the stretching of flexible and semi-flexible single polymer chains under external fields} 


\author{Fabio Manca} 
\affiliation{Department of Physics, University of Cagliari, 09042 Monserrato, Italy}
\author{Stefano Giordano} 
\email{stefano.giordano@iemn.univ-lille1.fr}
\affiliation{Institute of Electronics, Microelectronics and Nanotechnology (UMR CNRS 8520), 59652 Villeneuve d'Ascq, France} 
\affiliation{International Associated Laboratory LEMAC, ECLille, 59652 Villeneuve d'Ascq, France}
\author{Pier Luca Palla}
\affiliation{Institute of Electronics, Microelectronics and Nanotechnology (UMR CNRS 8520), 59652 Villeneuve d'Ascq, France}
\affiliation{University of Lille I, 59652 Villeneuve d'Ascq, France}  
\author{Fabrizio Cleri} 
\affiliation{Institute of Electronics, Microelectronics and Nanotechnology (UMR CNRS 8520), 59652 Villeneuve d'Ascq, France} 
\affiliation{University of Lille I, 59652 Villeneuve d'Ascq, France}
\author{Luciano Colombo}
\affiliation{Department of Physics, University of Cagliari, 09042 Monserrato, Italy}

\date{\today}

\begin{abstract}
Recent developments of microscopic mechanical experiments allow the manipulation of individual polymer molecules in two main ways: \textit{uniform stretching} by external forces and \textit{non-uniform stretching} by external fields. Many results can be thereby obtained for specific kinds of polymers and specific geometries. In this work we describe the non-uniform stretching of a single, non-branched polymer molecule by an external field (e.g. fluid in uniform motion, or uniform electric field) by a universal physical framework which leads to general conclusions on different types of polymers.  We derive analytical results both for the freely-jointed chain and the worm-like chain models based on classical statistical mechanics. Moreover, we provide a Monte Carlo numerical analysis of the mechanical properties of flexible and semi-flexible polymers anchored at one end. The simulations confirm the analytical achievements, and moreover allow to study the situations where the theory can not provide explicit and useful results. In all cases we evaluate the average conformation of the polymer and its fluctuation statistics as a function of the chain length, bending rigidity and field strength.
\end{abstract}

\pacs{}

\maketitle 



\section{Introduction}
Modern methods for stretching single molecules provide a valuable insight about the response of polymers to external forces. The interest on single molecules loading encouraged new research and technological developments on related mechanical experiments. Typically, mechanical methods allow the manipulation of a polymer molecule in two ways: the stretching of the chain by the direct action of an external force or by the application of an external field. If we consider homogeneous polymers (with all monomers described by the same effective elastic stiffness), then we obtain a uniform strain with the external force and a non-uniform strain with the applied field. 

To exert an external force on a polymer fixed at one end, laser optical tweezers (LOTs)\cite{lots}, magnetic tweezers (MTs)\cite{mts} or atomic force microscope (AFM)\cite{AFM} can be used. Many experiments have been performed over a wide class of polymers with biological relevance, such as the nucleic acids (DNA, RNA)\cite{nucleic}, allowing the stretching of the entire molecule and providing the reading and the mapping of genetic information along the chain.\cite{bensimon, chan} Furthermore, it has been possible to describe the elastic behaviour of single polymers consisting of domains which may exhibit transitions between different stable states.\cite{rief1, rief2, mancaII}
Other investigations performed on double-stranded DNA determined the extension of the polymer as a function of the applied force\cite{busta0}, providing results in very good agreement with the Worm-Like Chain (WLC) model\cite{smith1, marko,manca} and the Freely-Jointed Chain (FJC) model.\cite{manca,huguet} 

Alternatively, it is possible to manipulate single molecules by an external field. In this case the external field acts on the molecules from a distance or, in other words, without a defined contact point for applying the traction. A non-uniform stretching performed by an external field can be induced either via a hydrodynamic (or electrohydrodynamic) flow field\cite{trahan, wang, hsieh} or via an electric (or magnetic) field.\cite{schwartz, strick1, strick2} One experimental advantage of using flow fields is that the liquid surrounding the tethered molecule can be easily replaced; this is indeed an important feature for many single-molecules studies of enzymes which require varying buffer conditions.\cite{busta_cat} The flow field technique was extensively applied in single-molecule study of DNA elasticity\cite{smith1} as well as  to characterize the rheological properties of individual DNA molecules.\cite{smith2,perkins1,perkins2} The use of an electric field has been adopted for driving the alignment of DNA on a solid surface for applications such as gene mapping and restriction analysis.\cite{schwartz} Finally, magnetic fields have been used to apply torsional stress to individual DNA molecules.\cite{strick1,strick2}

In order to understand the response of polymers to external fields and to study their  statistics, some theoretical models have been proposed. These models are typically based on the FJC and WLC schemes, generalized with the inclusion of the given applied field. Some studies have shown that in a weak external field the persistence length along the field direction is increased, while it is decreased in the perpendicular direction; moreover, as the external field becomes stronger, the effective persistence length grows exponentially with the field strength.\cite{warner, vroege, kamien} Other investigations under a constant velocity flow have shown that a flexible polymer  displays three types of conformation: unperturbed at low velocity; ``trumpet'' shaped when partially stretched; ``stem and flowers'' shaped, with a completely stretched portion (the stem) and a series of blobs (the flowers), at larger loading.\cite{brochard1, brochard2, brochard3} Polymer models have been studied in elongational flows to analyze the coil stretching and chain retraction as a function of polymer and flow parameters, finding good agreement with experimental data.\cite{henvey, rabin} Conformational properties of semiflexible polymer chains in uniform force field were also studied for two-dimensional models.\cite{lamura}
In spite of all these relevant efforts, it is yet a challenge to base  on one same unified theoretical framework and understanding of all aspects of polymer mechanics in an external field.

Building on our previous studies,\cite{mancaII,manca} in this paper we study the conformational and mechanical properties of flexible and semi-flexible non-branched polymer model chains tethered at one end and immersed in an external force field. This situation is useful to describe almost two physical conditions of interest: a polymer chain immersed in a fluid in a uniform motion (our model is valid only when the action of the fluid motion can be described by a distribution of given forces applied to all monomers) and an arbitrarily charged chain inserted in a uniform electric field.

Our theoretical approach is twofold, since we adopt both analytical (statistical mechanics\cite{gibbs,weiner}) and numerical techniques (Monte Carlo simulations\cite{binder,confser}). While the analytical approach is useful to obtain the explicit partition function in some specific cases, Monte Carlo simulations are crucial to study more generic cases, inaccessible to analytical treatments. In particular, while we develop our theoretical framework starting from the more tractable FJC model, we take full profit from our MC simulations to extend our study also to the WLC model. 

The structure of the paper is the following. In Section II we introduce the mathematical formalism adopted and we derive a generic form of the partition function in $\Re^d$ for a generalized FJC model where the extensibility of the bonds is taken into account. In the Section III we find the two specific forms of the partition function for the 2D- and the 3D-case for the pure FJC polymer with non extensible bonds. Moreover, we obtain in both cases the variance and the covariance among the positions of the monomers. In the Section IV we present the generalization of previous results to the semi-flexible WLC model. We present two closed-forms approximations for the 2D- and the 3D-case and the comparisons with MC simulations. In section V we analyze the behavior of a chain in an external field to which also an external force is applied at the end of the chain. The case with the force not aligned with the field is particularly interesting and shows the power of the MC method. Finally, in Section VI some conclusions are drawn.

\section{General theoretical framework}
As previously discussed, the polymer models most used in literature are the FJC and the WLC. As argued in Ref.\onlinecite{cohen}, for weak tension and weak external field, it is acceptable to model the polymer as a FJC model. This model breaks down only when the curvature of the conformation is very large because it ignores the consequent great bending energy. Since we will look upon this problem in the end of this work, we now give way to the case of a FJC. In particular we consider a FJC with two additional hypothesis. Firstly we consider the possible extensibility of the bonds of the chain through a standard quadratic potential characterized by a given equilibrium length: such an extension mimics the possible stretching of the chemical bond between two adjacent monomers. If necessary, the extensibility of the bonds, here described by linear springs, can be easily extended to more complex, nonlinear springs.\cite{blundell} Moreover, we take into account a series of arbitrary forces applied to each monomer: these actions mimic the effects of an external physical field applied to the system. In addition, we contemplate the presence of an arbitrary force applied to the terminal monomer of the chain. 
All calculations will be performed in $\Re^d$ and we will specialize the results both in the 2D-case and in the 3D-case when needed. The idea is to write the complete form 
of the Hamiltonian of the system and to build up the corresponding statistical mechanics.\cite{manca} The starting point is therefore the calculation of the classical partition function. In fact, when this quantity is determined, it is possible to obtain the force-extension curve (the equation of state) through simple derivations. 

\begin{figure}
 \resizebox{1.0\columnwidth}{!}{\includegraphics{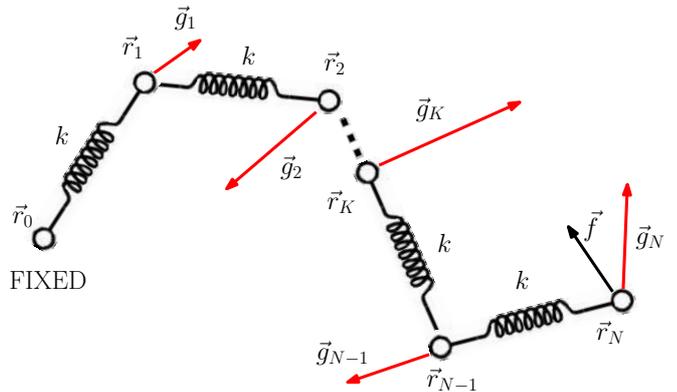}}
 \caption{(color online) A polymer chain in an external field. The first monomer is clamped at position $\vec r_0$ while the others are free to fluctuate. Each monomer is subjected to an external force $\vec g_K$ (different in strength and direction for any $K$): all these forces mimic an external field. Another external force, playing the role of a main pulling load, $\vec f$, is applied to the last monomer at the position $\vec r_N$.
 }
 \label{polymer_model}       
 \end{figure}

Let us consider a non-branched linear polymer with $N$ monomers (see Fig. \ref{polymer_model}) at positions defined by  $\vec r_1, ... ,\vec r_N \in \Re^d $ (for considering $d = 2$ or $d = 3$ according to the specific problem of interest). To each monomer a given external force is applied and named $\vec g_1, ... ,\vec g_N$. Another external force, playing the role of main pulling load, $\vec f$, is applied to the last monomer at the position $\vec r_N$. While the chain is clamped at position $\vec r_0$, the monomers are free to fluctuate. The Hamiltonian of the system is therefore given by 
\begin{eqnarray}
\label{hamiltonian}
H &=& \sum_{i=1}^{N} \frac{\vec p_i \cdot \vec p_i}{2m} +\frac{1}{2}k\sum_{K=1}^{N}\left( | \vec{r}_K-\vec{r}_{K-1}|-l\right)^{2} \\
\nonumber
&&- \sum_{K=1}^{N} \vec g_K \cdot \vec r_K - \vec f \cdot \vec r_N
\end{eqnarray}
where $\vec p_i$ are the linear momenta, $m$ the mass of the monomers, $k$ the spring constant of the inter-monomer interaction, and $l$ the equilibrium length of the monomer-monomer bond.
We search for the partition function of the system defined as:
\begin{eqnarray}
Z_d = c \underbrace{\int_{\Re^d} ... \int_{\Re^d}}_{2N- \mbox{times}} \exp \left( -\frac{H}{k_BT} \right) d\vec r_1 ... d\vec r_N d\vec p_1 ... d\vec p_N
\end{eqnarray}
where $c$ is a multiplicative constant which takes into account the number of microstates. As well known, the kinetic part can be straightforwardly integrated and it yields a further non-influencing multiplicative constant; then we can write the partition function as an integral over the positional space only.
This integral can be easily handled through the standard change of variable
\begin{eqnarray}
\left\{
\begin{array}{ll}
\vec \xi_1 = \vec r_1 - \vec r_0 \\
\vec \xi_2 = \vec r_2 - \vec r_1 \\
\hspace{0.6cm} \vdots \\
\vec \xi_N = \vec r_N - \vec r_{N-1} \\
\end{array}
\right.
\end{eqnarray}
having the Jacobian determinant $J = \left| \frac{\partial(\vec r_1 ... \vec r_N)}{\partial(\vec \xi_1 ... \vec \xi_N)} \right| = 1$. We consider the terminal $\vec r_0$ of the chain fixed in the origin of axes, i.e. $\vec r_0 = \vec 0$.
So, we cast the positions $\vec r_i$ in terms of the variables $\vec \xi_J$ as follows
\begin{eqnarray}
\left\{
\begin{array}{ll}
\vec r_1 = \vec \xi_1 + \vec r_0 = \vec \xi_1 \\
\vec r_2 = \vec \xi_2 + \vec r_1 = \vec \xi_2 + \vec \xi_1 \\
\hspace{0.6cm} \vdots \\
\vec r_N = \vec \xi_N + \vec \xi_{N-1} + ... + \vec \xi_1 \\
\end{array}
\right.
\end{eqnarray}
By setting the general solution as $ \vec r_i = \sum_{K=1}^{i} \vec \xi_K $, the partition function becomes
\begin{eqnarray}
Z_d &=& c \underbrace{\int_{\Re^d} ... \int_{\Re^d}}_{N- \mbox{times}} \exp \left[  -\frac{k}{2k_BT}\sum_{K=1}^{N}\left( |\vec{\xi}_K|-l\right)^{2}
 \right] \\
\nonumber
&& \times \exp \left[ \frac{1}{k_BT}\sum_{K=1}^{N} \vec g_K \cdot \sum_{J=1}^{K} \vec \xi_J \right]\\
\nonumber
&& \times \exp \left[ \frac{1}{k_BT} \vec f \cdot \sum_{K=1}^{N} \vec \xi_K \right]  d\vec \xi_1 ... d\vec \xi_N
\end{eqnarray}
Inverting the two summation symbols
\begin{eqnarray}
\sum_{K=1}^{N} \vec g_K \cdot \sum_{J=1}^K \vec \xi_J = \sum_{K=1}^{N} \vec \xi_K \cdot \sum_{i=K}^{N} \vec g_i 
\end{eqnarray}
we obtain
\begin{eqnarray}
\label{partition1}
Z_d = c \prod_{K=1}^{N} \int_{\Re^d} e^{ -a \left( |\vec{\xi}|-l\right)^{2} } e^{ \vec V_K \cdot \vec \xi } d\vec \xi
\end{eqnarray}
where
\begin{eqnarray}
  \label{partval1}
  a &=& \frac{k}{2k_BT} >  0
\\
\label{partval2}
  \vec V_K &=& \frac{1}{k_BT} \left( \vec f + \sum_{i=K}^{N} \vec g_i \right)
\end{eqnarray}
It exists a deep conceptual connection between the last integral for the partition function and the theory of the $d$-dimensional Fourier transforms. The Fourier integral of an arbitrary function $f(\vec \xi)$ is defined as
\begin{equation}
F(\vec \omega) = \int_{\Re^d} f(\vec \xi) e^{-i \vec \omega \cdot \vec \xi} d\vec \xi
\end{equation}
with inverse transform given by
\begin{equation}
 f(\vec \xi) = \frac{1}{(2 \pi)^d} \int_{\Re^d} F(\vec \omega) e^{i \vec \omega \cdot \vec \xi} d\vec \omega
\end{equation}
If we consider 
\begin{equation}
\label{function}
 f(\vec \xi) = e^{ -a \left( |\vec{\xi}|-l\right)^{2} }
\end{equation}
 it is easy to realize that the integral in Eq.(\ref{partition1}) is the Fourier transform of $f(\vec \xi)$ calculated for $\vec \omega = i \vec V_K$, i.e.
\begin{equation}
\label{partFourier}
 Z_d = c \prod_{K=1}^{N} F(i \vec V_K)
\end{equation}
with $a$ e $\vec V_K$ defined respectively in Eq.(\ref{partval1}) and Eq.(\ref{partval2}).
It is important to remark that the function in Eq.(\ref{function}) has a spherical symmetry (i.e. it depends only on the length of the vector $\vec \xi$) and, therefore, also its Fourier transform $F(\vec \omega)$ exhibits the spherical symmetry, depending only on the quantity $|\vec \omega|$ in the transformed domain. In fact, for such spherically-symmetric functions it holds that: if $f(\vec \xi) = f(|\vec \xi|) $ then $ F(\vec \omega) = F(|\vec \omega|) $. Furthermore, we have that
\begin{equation}
 F(\Omega) = \int_{0}^{+\infty} 2\pi\rho f(\rho) \left( \frac{2\pi\rho}{\Omega} \right)^{\frac{d}{2}-1} J_{\frac{d}{2}-1}(\rho\Omega) d\rho    
\end{equation}
for $ d=2n  \mbox{ (even)}$, and
\begin{equation}
 F(\Omega) = \int_{0}^{+\infty} 4\pi\rho^2 f(\rho) \left( \frac{2\pi\rho}{\Omega} \right)^{\frac{d-3}{2}} j_{\frac{d-3}{2}}(\rho\Omega) d\rho   
\end{equation}
for $ d=2n+1  \mbox{ (odd)} $, where $\rho = |\vec \xi|$ and $\Omega = |\vec \omega|$.\cite{schwartz-math} Here $J_{\nu}(z)$ and $j_{\nu}(z)$ are the cylindrical and spherical Bessel functions of the first kind respectively, correlated by the standard relation $j_{\nu}(z) = \sqrt{\frac{\pi}{2z}}J_{{\nu}+\frac{1}{2}}(z)$.\cite{abra,grad}
In our calculations we have to set $\vec \omega = i \vec V_K$ and, therefore, we obtain $ \Omega =  i |\vec V_K|$.
Moreover, when the argument of $J_{\nu}(z)$ and $j_{\nu}(z)$ is supposed imaginary we obtain the modified Bessel functions of the first kind\cite{abra,grad}
\begin{equation}
\begin{array}{ll}
I_{\nu}(z) = (i)^{-\nu} J_{\nu}(iz) \\
i_{\nu}(z) = (i)^{-\nu} j_{\nu}(iz) \\
\end{array}
\end{equation}
For example we have the explicit expression $j_0(z)=\frac{\sin z}{z}$ and $i_0(z)=\frac{\sinh z}{z}$ while, on the contrary,  $I_0(z)$ and $J_0(z)$ cannot be written in closed form.
So, for $d$ even we eventually obtain
\begin{equation}
 \frac{F(i \vec V_K)}{2\pi} = \int_{0}^{+\infty} \rho \hspace{0.1cm} e^{-a(\rho-l)^2} \left( \frac{2\pi\rho}{|\vec V_K|} \right)^{\frac{d-2}{2}} I_{\frac{d-2}{2}}(\rho|\vec V_K|) d\rho   
\end{equation}
and, on the other hand, for $d$ odd we have
\begin{equation}
\frac{F(i \vec V_K)}{4\pi}  = \int_{0}^{+\infty} \rho^2 \hspace{0.1cm} e^{-a(\rho-l)^2} \left( \frac{2\pi\rho}{|\vec V_K|} \right)^{\frac{d-3}{2}} i_{\frac{d-3}{2}}(\rho|\vec V_K|) d\rho   
\end{equation}
Finally, by using Eq.(\ref{partFourier}), the partition function is given by
\begin{equation}
\label{partition_even}
Z_d = c \prod_{K=1}^{N} \int_{0}^{+\infty} \rho \hspace{0.1cm} e^{-a(\rho-l)^2} \left( \frac{\rho}{|\vec V_K|} \right)^{\frac{d-2}{2}} I_{\frac{d-2}{2}}(\rho|\vec V_K|) d\rho  
\end{equation}
for $d$ even, and
\begin{equation}
\label{partition_odd}
Z_d = c \prod_{K=1}^{N} \int_{0}^{+\infty} \rho^2 \hspace{0.1cm} e^{-a(\rho-l)^2} \left( \frac{\rho}{|\vec V_K|} \right)^{\frac{d-3}{2}} i_{\frac{d-3}{2}}(\rho|\vec V_K|) d\rho  
\end{equation}
for  $d$ odd, where $ a$  and $ \vec V_K $ are given in
Eqs.(\ref{partval1}) and (\ref{partval2}). In the framework of statistical mechanics, the knowledge of the partition function allows to determine all needed expected values describing the statistics of the chain (i.e., average values of the positions, variances of the positions and so on).

\section{Freely-jointed chain model under external field}

\subsection{Average values of positions}
\begin{figure}[ht]
 \resizebox{0.7\columnwidth}{!}{\includegraphics{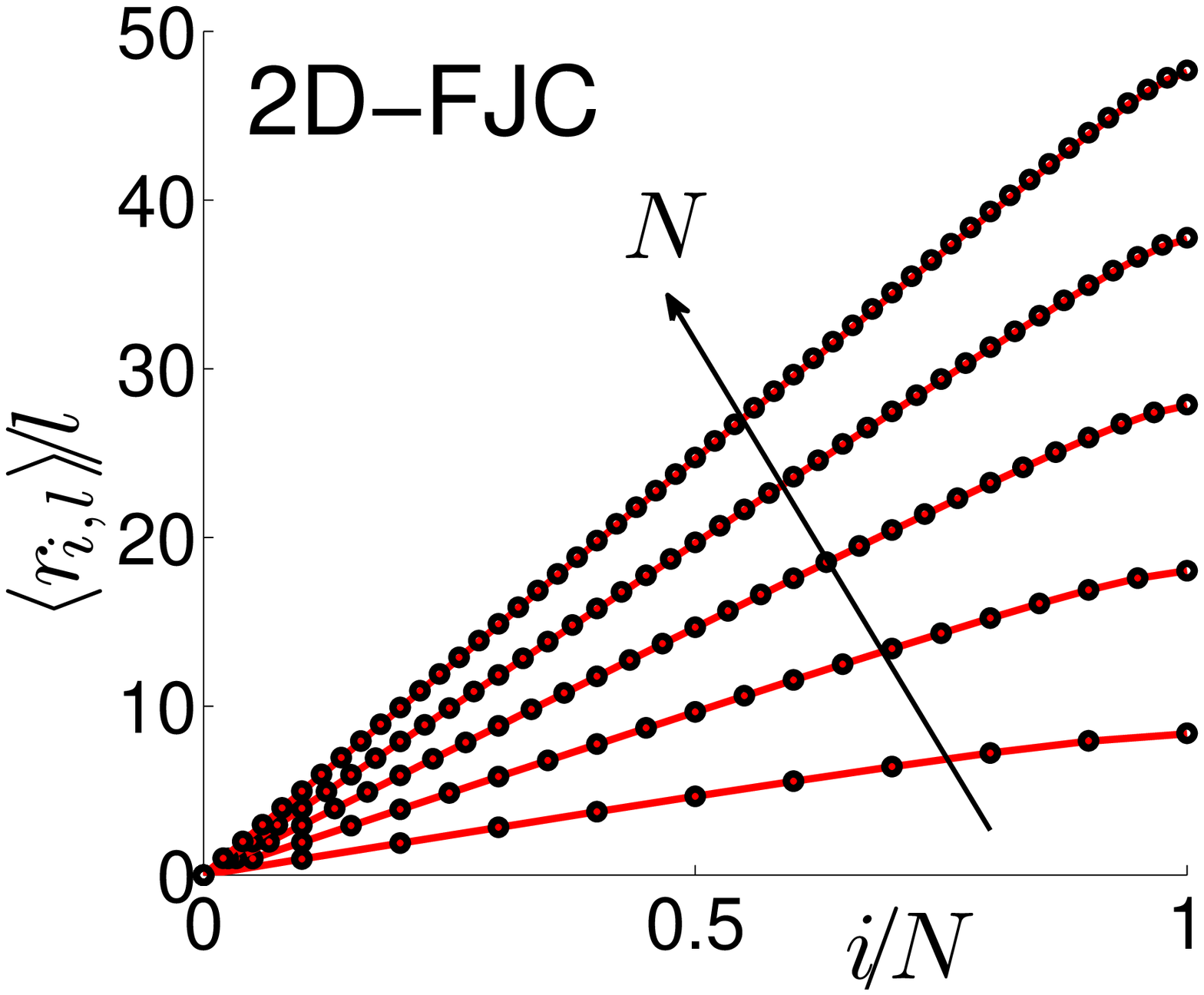}}\\
 \resizebox{0.7\columnwidth}{!}{\includegraphics{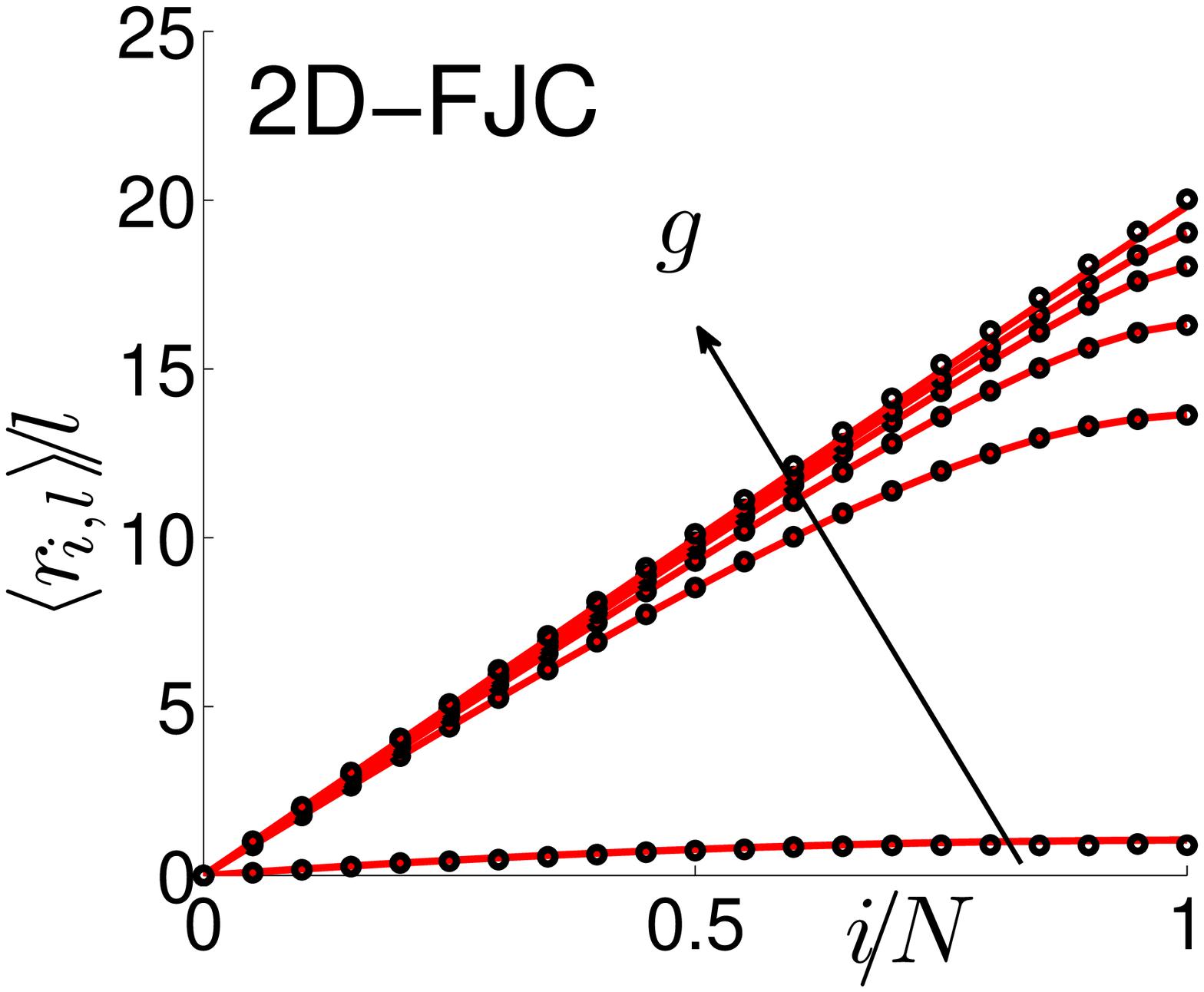}}
 \caption{(color online) Average values of the longitudinal component of the positions induced by the external field for the 2D FJC case. The red solid lines correspond to the analytical results Eqs.(\ref{2Dcampo}) and (\ref{fjc2Dg}), MC results are superimposed in black circles. Top panel: each curve corresponds to different chain lengths $N=10, 20, 30, 40 ,50$ for a fixed value ${gl}/(k_B T)=1$ (e.g., corresponding to $l=1$nm, $g=4$pN at $T=293$K). Bottom panel: each curve corresponds to the different values  $gl/(k_B T)=0.1, 0.25, 0.5, 1, 2, 10$ for a fixed chain length $N=20$.
 }
 \label{positions_2D}       
 \end{figure}

In the previous section we obtained the general expression of the partition function for the case where the extensibility of the bonds is taken into account. This is described by the parameter $k$, which characterizes the elastic bond between adjacent monomers. In the present Section we want to study the effects of an arbitrary distribution of forces on a pure freely jointed chain model (FJC). Therefore  we need to obtain the specific form of the partition function in the case of rigid bonds of fixed length $l$.  From the mathematical point of view it means that we will consider $k \rightarrow \infty$, a condition representing a inextensible spring.
Because of the relation $\sqrt{\frac{\alpha}{\pi}} e^{-\alpha x^2} = \delta(x)$ when $\alpha \rightarrow \infty$ we may determine the limit of Eq.(\ref{partition_even}) and Eq.(\ref{partition_odd}) for $a \rightarrow \infty$ (i.e. for $k \rightarrow \infty$, FJC limit). Since the arbitrariness of the constant $c$, we may consider in Eqs. (\ref{partition_even}) and (\ref{partition_odd}) a multiplicative constant term $(\sqrt{\frac{a}{\pi}})^N$. Then, by using the translated property $\sqrt{\frac{a}{\pi}} e^{-a(\rho-l)^2} \rightarrow \delta(\rho-l)$ for $a \rightarrow \infty$ we perform all the integrals thereby obtaining
\begin{equation}
\label{partition_even2}
Z_d = c \prod_{K=1}^{N} \frac{1}{|\vec V_K|^{\frac{d-2}{2}}} I_{\frac{d-2}{2}}(l|\vec V_K|) \hspace{1cm} \mbox{$d$ even}
\end{equation}
\begin{equation}
\label{partition_odd2}
Z_d = c \prod_{K=1}^{N} \frac{1}{|\vec V_K|^{\frac{d-3}{2}}} i_{\frac{d-3}{2}}(l|\vec V_K|) \hspace{1cm} \mbox{$d$ odd} 
\end{equation}
In particular, for $d = 2$ we have 
\begin{equation}
\label{partition_d2}
Z_2 = c \prod_{K=1}^{N} I_0 \left( \frac{l}{k_BT} \left| \vec f + \sum_{i=K}^{N} \vec g_i \right|  \right)
\end{equation}
while for $d=3$ we obtain
\begin{equation}
\label{partition_d3}
Z_3 = c \prod_{K=1}^{N} \frac{\sinh \left( \frac{l}{k_BT} \left| \vec f + \sum_{i=K}^{N} \vec g_i \right|  \right)}{\frac{l}{k_BT} \left| \vec f + \sum_{i=K}^{N} \vec g_i \right|} 
\end{equation}
All the expressions given in Eqs.(\ref{partition_even2}), (\ref{partition_odd2}), (\ref{partition_d2}), (\ref{partition_d3}) can be summarized in the general form 
\begin{equation}
\label{partition_general}
Z_d = c \prod_{K=1}^{N} f(|\vec V_K|) 
\end{equation}
with a suitable function $f(x)$. By using this expression of the partition function we can find the average position of the $i$-th monomer of the chain; indeed, from the definition of the Hamiltonian in Eq.(\ref{hamiltonian}) we state that $\vec r_i = -\frac{\partial H}{\partial \vec g_i}$ and, therefore, we get
\begin{equation}
\label{shape}
\langle \vec r_i\rangle  = k_BT \frac{\partial}{\partial \vec g_i} \ln Z_d
\end{equation}
which represents the shape of the polymer chain under the effects of the external field $\vec g_i$ and the applied force $\vec f$. Now we can substitute Eq.(\ref{partition_general}) into Eq.(\ref{shape}), obtaining
\begin{equation}
  \label{shape3}
  \langle \vec r_i\rangle  = \sum_{K=1}^{i} \frac{\vec V_K}{|\vec V_K|} \left[ \frac{1}{f(x)} \frac{\partial f(x)}{\partial x} \right]_{x=|\vec V_K|} 
\end{equation}
In 2D we have $f(x) = I_0(lx)$ and therefore we obtain
\begin{equation}
\label{2Dcampo}
 \langle \vec r_i\rangle  = l \sum_{K=1}^{i} \frac{I_1 \left( \frac{l}{k_BT} \left| \vec f + \sum_{J=K}^{N} \vec g_J \right| \right)}{I_0 \left( \frac{l}{k_BT} \left| \vec f + \sum_{J=K}^{N} \vec g_J \right| \right)} \frac{\vec f + \sum_{J=K}^{N} \vec g_J}{\left| \vec f + \sum_{J=K}^{N} \vec g_J \right|}
\end{equation}
For such a 2D case, by applying Eq.(\ref{2Dcampo}), the average values of the longitudinal component of the positions have been calculated and are plotted in Fig.\ref{positions_2D} as a function of the chain length $N$ and the field strength $g$. We have considered only the action of an external uniform field with  $\vec g_J=\vec g$ and amplitude $g$.

Although this case lends itself to a full analytical solution, numerical simulations were also performed by using a conventional implementation of the Metropolis version of the Monte Carlo algorithm.\cite{binder} The initial state of the chain is defined by a set of randomly chosen positions. 
The displacement extent of each step governs the efficiency of the configurational space sampling.  Therefore, we analysed several runs in order to optimize its value.\cite{frenkel,allen}
The perfect agreement between the theory and the MC simulations provides a strict check of the numerical procedure, to be used in the foregoing.

\begin{figure}[ht]
 \resizebox{0.7\columnwidth}{!}{\includegraphics{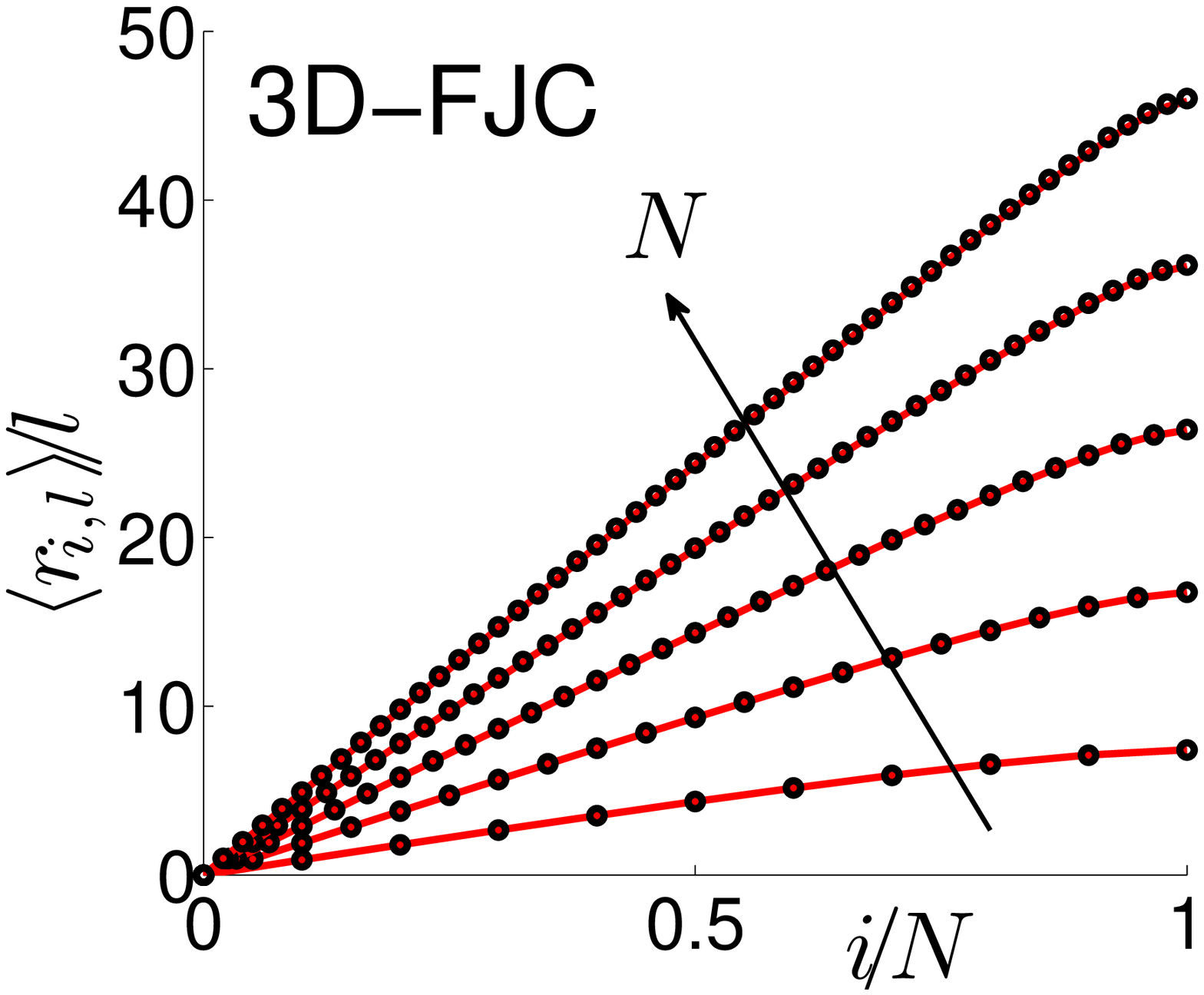}}\\
 \resizebox{0.7\columnwidth}{!}{\includegraphics{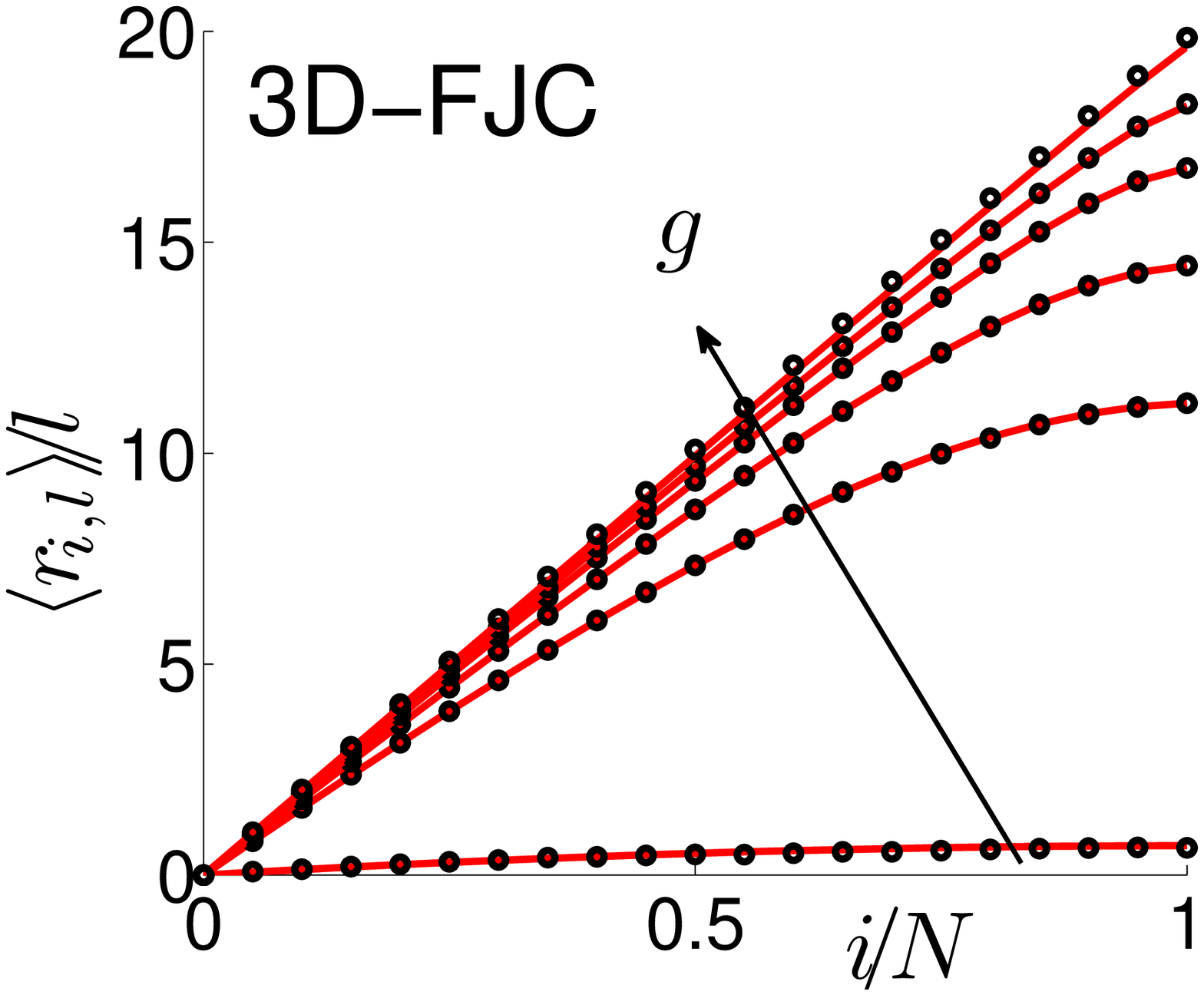}}
 \caption{(color online) Average values of the longitudinal component of the positions induced by the external field for the 3D FJC case. The red solid lines correspond to the analytical results Eqs.(\ref{3Dcampo}) and (\ref{fjc3Dg}), MC results are superimposed in black circles. Top panel: each curve corresponds to different chain lengths  $N=10, 20, 30, 40 ,50$ for a fixed value ${gl}/(k_B T)=1$. Bottom panel: each curve corresponds to the different values $gl/(k_B T)=0.1, 0.25, 0.5, 1, 2, 10$ for a fixed chain length $N=20$.
 }
 \label{positions_3D}       
 \end{figure}

On the other hand, in 3D we have $ f(x) = \frac{\sinh (lx)}{lx} $, leading to
\begin{equation}
\label{3Dcampo}
 \langle \vec r_i\rangle  = l \sum_{K=1}^{i} \mathcal{L} \left( \frac{l}{k_BT} \left| \vec f + \sum_{J=K}^{N} \vec g_J \right| \right)  \frac{\vec f + \sum_{J=K}^{N} \vec g_J}{\left| \vec f + \sum_{J=K}^{N} \vec g_J \right|}
\end{equation}
where $\mathcal{L}(x) = \coth x - \frac{1}{x}$ is the Langevin function. By using Eq.(\ref{3Dcampo}), as before, it is possible to plot the average values of the longitudinal component of the positions for the 3D case (Fig.\ref{positions_3D}).
Also in this case we adopted a uniform field $g$ and the good agreement with the MC simulations is evident.

 
As particular case, if there is only the force $\vec{f}$ applied to the system we obtain the standard scalar force-extension curves linking $r=\vert \langle\vec{r}_N\rangle\vert$ with $f=\vert \vec{f}\vert$. In 2D we have
\begin{equation}
\label{fjc2Df}
\frac{r}{lN}=\frac{I_1\left(\frac{lf}{k_BT} \right) }{I_0\left(\frac{lf}{k_BT} \right)}
\end{equation}
in agreement with recent results,\cite{kierfeld} while in 3D we obtain
\begin{equation}
\label{fjc3Df}
\frac{r}{lN}=\mathcal{L}\left(\frac{lf}{k_BT} \right) 
\end{equation}
which is a classical result.\cite{manca,rubinstein} The simple results in Eqs.(\ref{fjc2Df}) and (\ref{fjc3Df}) have been used to obtain the limiting behaviors under low ($f\rightarrow 0$) and high ($f\rightarrow \infty$) values of the applied force, as shown in Table \ref{asym}.

Building on such first results we now focus on some particular interesting approximations.
More specifically, it can be interesting to find approximate results for the case of a homogeneous field and no end-force, $\vec{f}=0$ and $\vec g_J=\vec g$ for any $J$. In this case we search for the scalar relation between
$r=\vert \langle\vec{r}_N\rangle\vert$ and $g=\vert \vec{g}\vert$. In the 2D case, from Eq.(\ref{2Dcampo}), we have
 \begin{eqnarray}
 \nonumber
 \frac{r}{lN} &=& \frac{1}{N} \sum_{k=1}^{N} \frac{I_1 \left( \frac{lg}{k_BT} (N-k+1) \right)}{I_0 \left( \frac{lg}{k_BT} (N-k+1) \right)} \\
 \nonumber
 &\simeq & \frac{1}{N} \int_{0}^{N} \frac{I_1 \left( \frac{lg}{k_BT} (N-x+1) \right)}{I_0 \left( \frac{lg}{k_BT} (N-x+1) \right)}dx \\
 &= & \frac{1}{N}\frac{1}{\frac{lg}{k_BT}} \log \frac{I_0 \left( \frac{lg}{k_BT} (N+1) \right)}{I_0 \left( \frac{lg}{k_BT} \right)} 
 \label{fjc2Dg}
\end{eqnarray}
On the other hand, for the 3D case we obtain
\begin{eqnarray}
\nonumber
 \frac{r}{lN}  &=& \frac{1}{N} \sum_{K=1}^{N} \mathcal{L} \left( \frac{l}{k_BT}(N-k+1) \right) \\
 \nonumber
 &\simeq & \frac{1}{N} \int_{0}^{N}\mathcal{L} \left( \frac{l}{k_BT}(N-x+1) \right)dx\\
 &= &\frac{1}{N}\dfrac{1}{\frac{lg}{k_BT}}\log \dfrac{\mbox{e}^{ 2\frac{lg}{k_BT} (N+1)}-1}{(N+1) \left(\mbox{e}^{2 \frac{lg}{k_BT} }-1\right)} -1
 \label{fjc3Dg}
\end{eqnarray}
We have usefully exploited the fact that, for large $N$, the sums can be approximately substituted with the corresponding integrals, which are easier to be handled. The closed-form expressions given in Eqs.(\ref{fjc2Dg}) and (\ref{fjc3Dg}) are very useful to obtain the limiting behaviors of the polymer under low ($g\rightarrow 0$) and high ($g\rightarrow \infty$) values of the applied field, as shown in Table \ref{asym}. Moreover, we have verified the validity of Eqs.(\ref{fjc2Dg}) and (\ref{fjc3Dg}) through a series of comparisons with MC results (see Fig.\ref{forcextension_FJC} in the next Section for details).

\subsection{Covariances and variances of positions}

In this Section, we search for the covariance among the positions of the monomers. It is important to evaluate such a quantity in order to estimate the variance of a given position (measuring the width of the probability density around its average value) and the correlation among different monomer positions (measuring the persistence of some geometrical features along the chain). In order to do this, we identify the $\alpha$-th component of the $i$-th monomer as $r_{i \alpha}$. The covariance of the generic monomer simply defined as (it represent the expectation value of the second order):
\begin{eqnarray}
 \mbox{Cov}(r_{i \alpha}, r_{J \beta}) &=& \langle  (r_{i \alpha} - \langle r_{i \alpha}\rangle )(r_{J \beta} - \langle r_{J \beta}\rangle ) \rangle   \\ \nonumber
&=& \langle r_{i \alpha} r_{J \beta}\rangle  - \langle r_{i \alpha}\rangle  \langle r_{J \beta}\rangle 
\end{eqnarray}
Taking the derivative of the partition function with respect to the $\alpha$ and the $\beta$ components of the force vectors $\vec g_i$ and $\vec g_J$ we can solve the problem as follows. We consider the standard expression for the partition function
and we can elaborate the following expression
\begin{eqnarray}
 \label{correlation_term}
 \langle r_{i \alpha} r_{J \beta}\rangle  = (k_BT)^2 \left( \frac{\partial \ln Z_d}{\partial g_{i \alpha}} \frac{\partial \ln Z_d}{\partial  g_{J \beta}}   + \frac{\partial^2 \ln Z_d}{\partial g_{i \alpha} \partial  g_{J \beta} } \right)
 \end{eqnarray}
or, equivalently, by introducing Eq.(\ref{shape})
\begin{eqnarray}
 \langle r_{i \alpha} r_{J \beta}\rangle = \langle r_{i \alpha}\rangle  \langle r_{J \beta}\rangle  + k_BT \frac{\partial}{\partial  g_{J \beta}} \langle r_{i \alpha} \rangle\,\,\,\,
 \end{eqnarray}
but we can simply determine that
 \begin{eqnarray}
 \frac{\partial}{\partial  g_{J \beta}} \langle r_{i \alpha}\rangle  = \frac{\partial}{\partial  g_{J \beta}} \sum_{K=1}^{i} \frac{\vec V_K \cdot \vec e_\alpha}{|\vec V_K|} \left[ \frac{1}{f(x)} \frac{\partial f(x)}{\partial x} \right]_{x=|\vec V_K|} 
 \end{eqnarray}
where we have defined the unit vector $\vec e_{\alpha}$ as the basis of the orthonormal reference frame. Being
 \begin{eqnarray} 
 \vec V_K \cdot \vec e_\alpha = \frac{1}{k_BT} \left( f_\alpha + \sum_{i=K}^{N} g_{i \alpha} \right) 
 \end{eqnarray}
 and 
  \begin{eqnarray} \frac{\partial |\vec V_K|}{\partial g_{J \beta}} = \frac{1}{k_BT} \frac{\vec V_K \cdot \vec e_\beta}{|V_K|} \sum_{q=K}^{N} \delta_{Jq} 
  \end{eqnarray}
after long but straightforward calculations we obtain
 \begin{eqnarray}
&& k_BT \frac{\partial}{\partial  g_{J \beta}} \langle r_{i \alpha}\rangle  = \sum_{K=1}^{\mbox{min}\{i,J\}} \frac{1}{|\vec V_K|f(|\vec V_K|)}  \\ 
\nonumber
&&\times \left\{ \delta_{\alpha \beta} f'(|\vec V_K|) + f''(|\vec V_K|) \frac{V_{K\alpha} V_{K\beta}}{|\vec V_K|}  \right. \\ \nonumber
&&- \left. V_{K\alpha} f'(|\vec V_K|) \frac{V_{K\beta}}{|\vec V_K|^2} - V_{K\alpha} 
 \frac{ f'(|\vec V_K|)^2}{f(|\vec V_K|)}  \frac{V_{K\beta}}{|\vec V_K|} \right\} 
 \end{eqnarray}
Ordering the terms we finally obtain the important result 
 \begin{eqnarray}
\mbox{Cov}(r_{i \alpha}, r_{J \beta})
 &=& \sum_{K=1}^{\mbox{min}\{i,J\}} \frac{\delta_{\alpha \beta}}{|\vec V_K|} \frac{f'(|\vec V_K|)}{f(|\vec V_K|)} \\
 \nonumber &+& \sum_{K=1}^{\mbox{min}\{i,J\}} \frac{V_{K\alpha} V_{K\beta} }{|\vec V_K|^2 f(|\vec V_K|)} \\ \nonumber
 &\times& \left \{  f''(|\vec V_K|) - \frac{f'(|\vec V_K|)}{|\vec V_K|}  - \frac{ f'(|\vec V_K|)^2}{f(|\vec V_K|)} \right\}
 \end{eqnarray}
It represents the final form of the covariance between two different components of the positions of two different monomers.

If we look at the variance of a single component of a single position ($i=J$, $\alpha = \beta$) we have the simpler result
 \begin{eqnarray}
 \label{variance}
\sigma^{2}_{i \alpha}
 &=& \sum_{K=1}^{i} \frac{f'(|\vec V_K|)}{|\vec V_K|f(|\vec V_K|)}  + \sum_{K=1}^{i} \frac{V_{K\alpha}^2 }{|\vec V_K|^2 f(|\vec V_K|)} \\ \nonumber
 &\times& \left \{  f''(|\vec V_K|) - \frac{f'(|\vec V_K|)}{|\vec V_K|}  - \frac{ f'(|\vec V_K|)^2}{f(|\vec V_K|)} \right\}
 \end{eqnarray}
In order to use the previous expressions we have to specify the function $f$ and its derivatives for the two-dimensional and the three-dimensional case.
In the 2D case we have $ f(x) = I_0(lx)$, $  f'(x) = l I_1(lx) $ and $ f''(x) = \frac{l^2}{2} [I_0(lx) + I_2(lx)] $. On the other hand, for the 3D case we have
$ f(x) = \frac{\sinh(lx)}{lx} $, $ f'(x)/f(x) = l \mathcal{L}(lx)$ and $ f''(x)/f(x) = l^2 -2l \mathcal{L}(lx) /x$.
This completes the determination of the covariance.

\begin{figure}[ht]
 \hspace{-0.7cm}\resizebox{0.7\columnwidth}{!}{\includegraphics{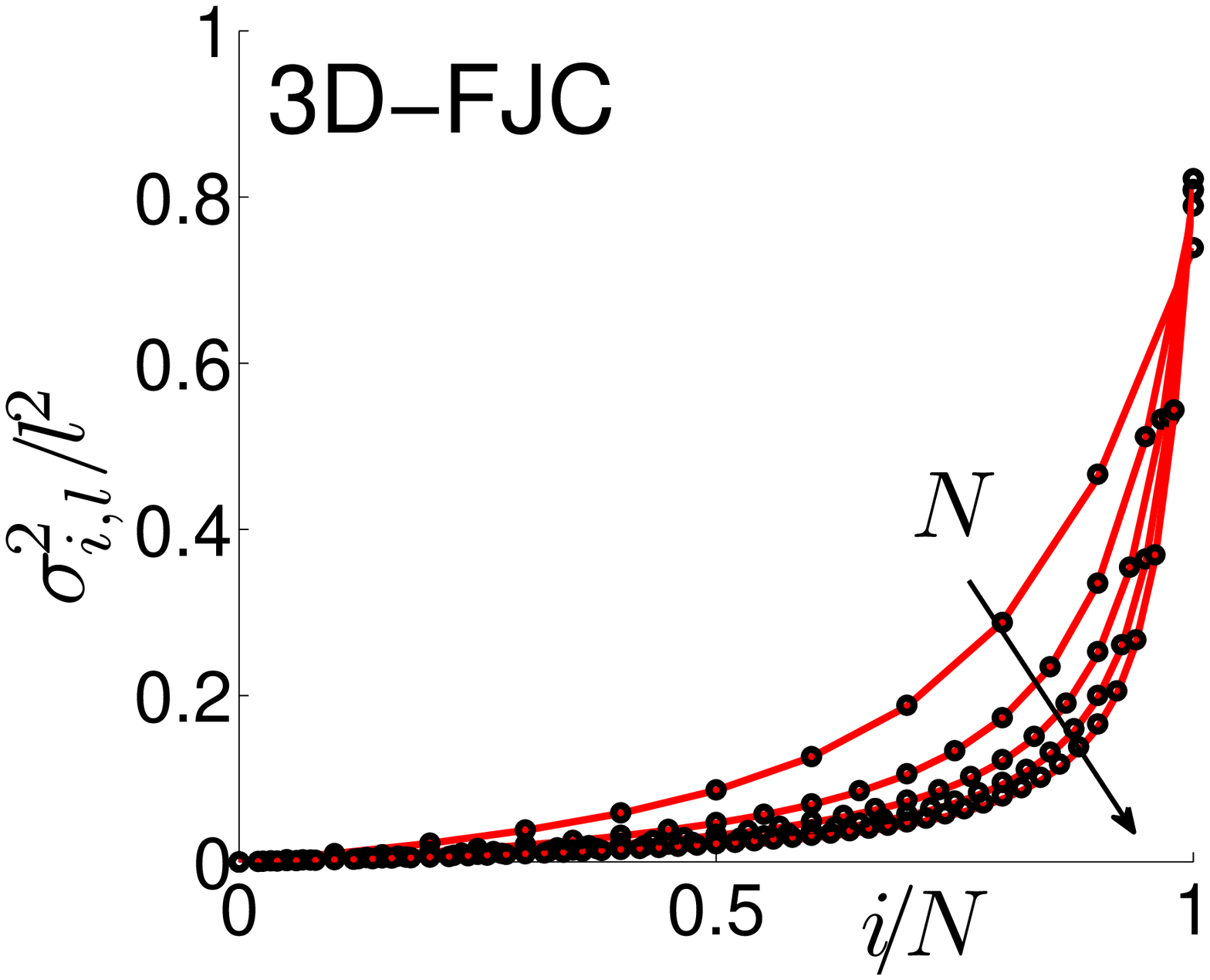}}\\
 \resizebox{0.7\columnwidth}{!}{\includegraphics{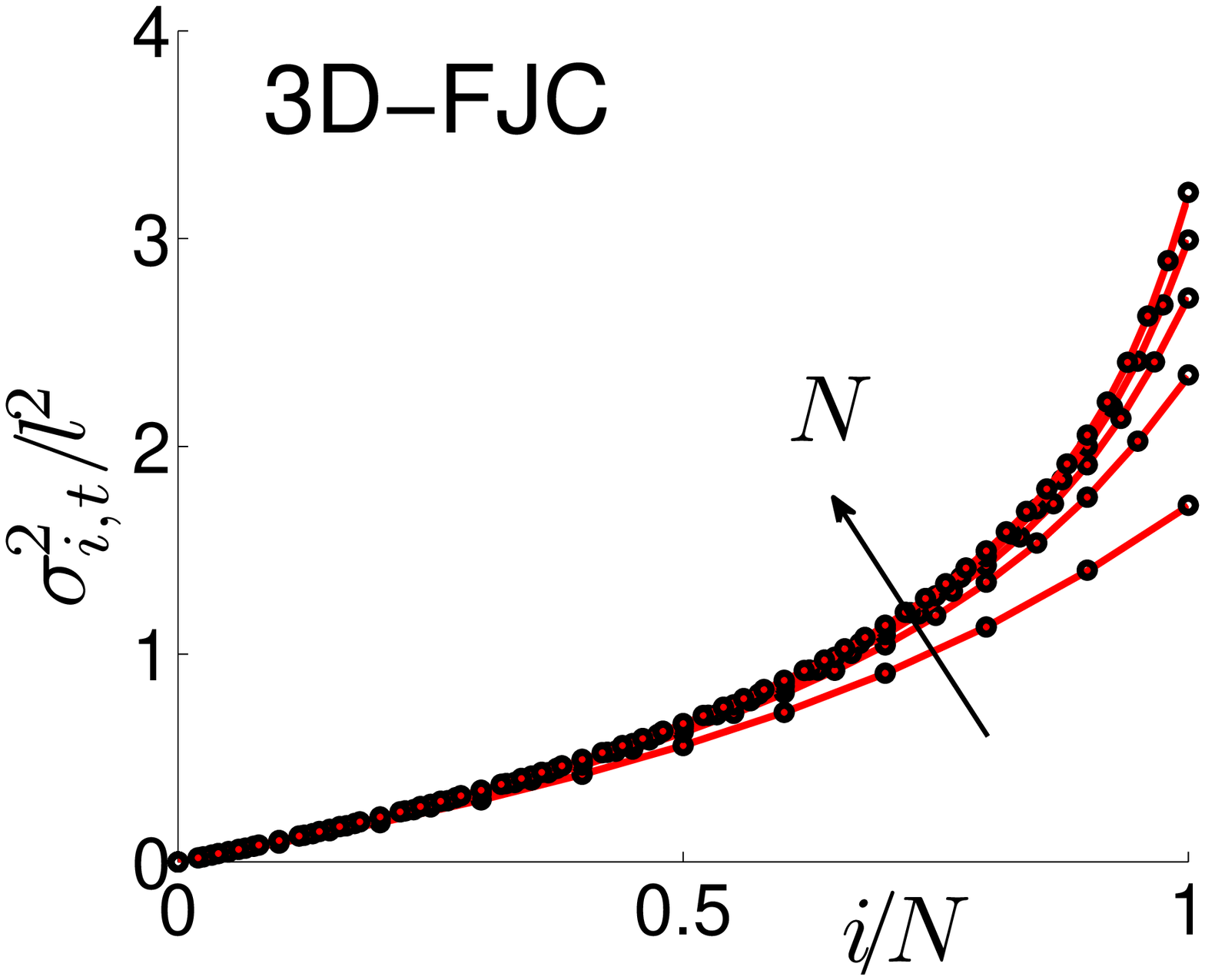} }
 \caption{(color online) Longitudinal (top panel) and transversal (bottom panel) component of the variance of positions for the 3D FJC case. The red solid lines correspond to the analytical result Eq.(\ref{variance}), MC results are superimposed in black circles. Each curve corresponds to different chain lengths  $N=10, 20, 30, 40 ,50$ for a fixed value of the external field defined by ${gl}/(k_B T)=1$.
 }
 \label{variances_3D_N}       
 \end{figure}

\begin{figure}[ht]
 \resizebox{0.7\columnwidth}{!}{\includegraphics{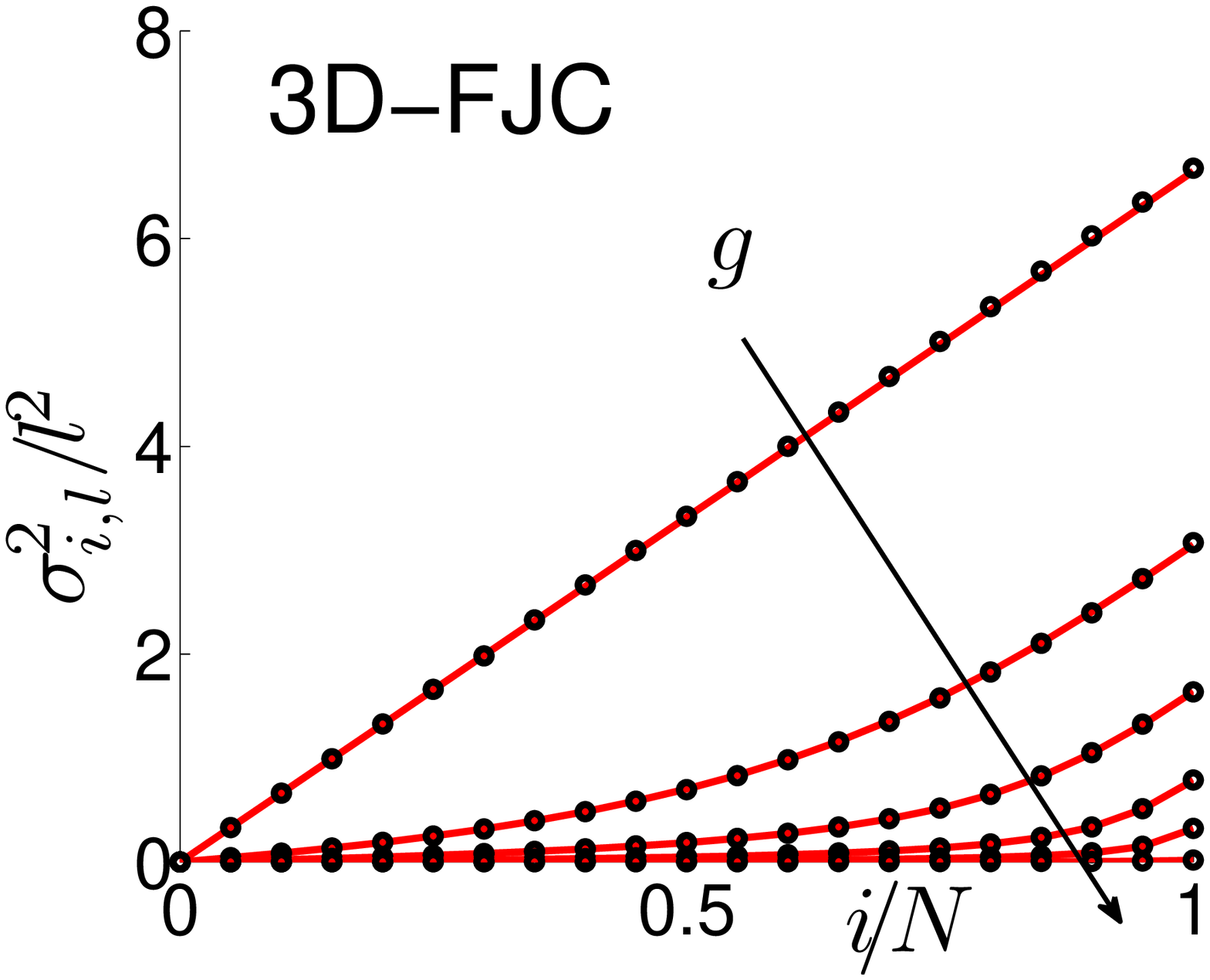}}\\
 \resizebox{0.7\columnwidth}{!}{\includegraphics{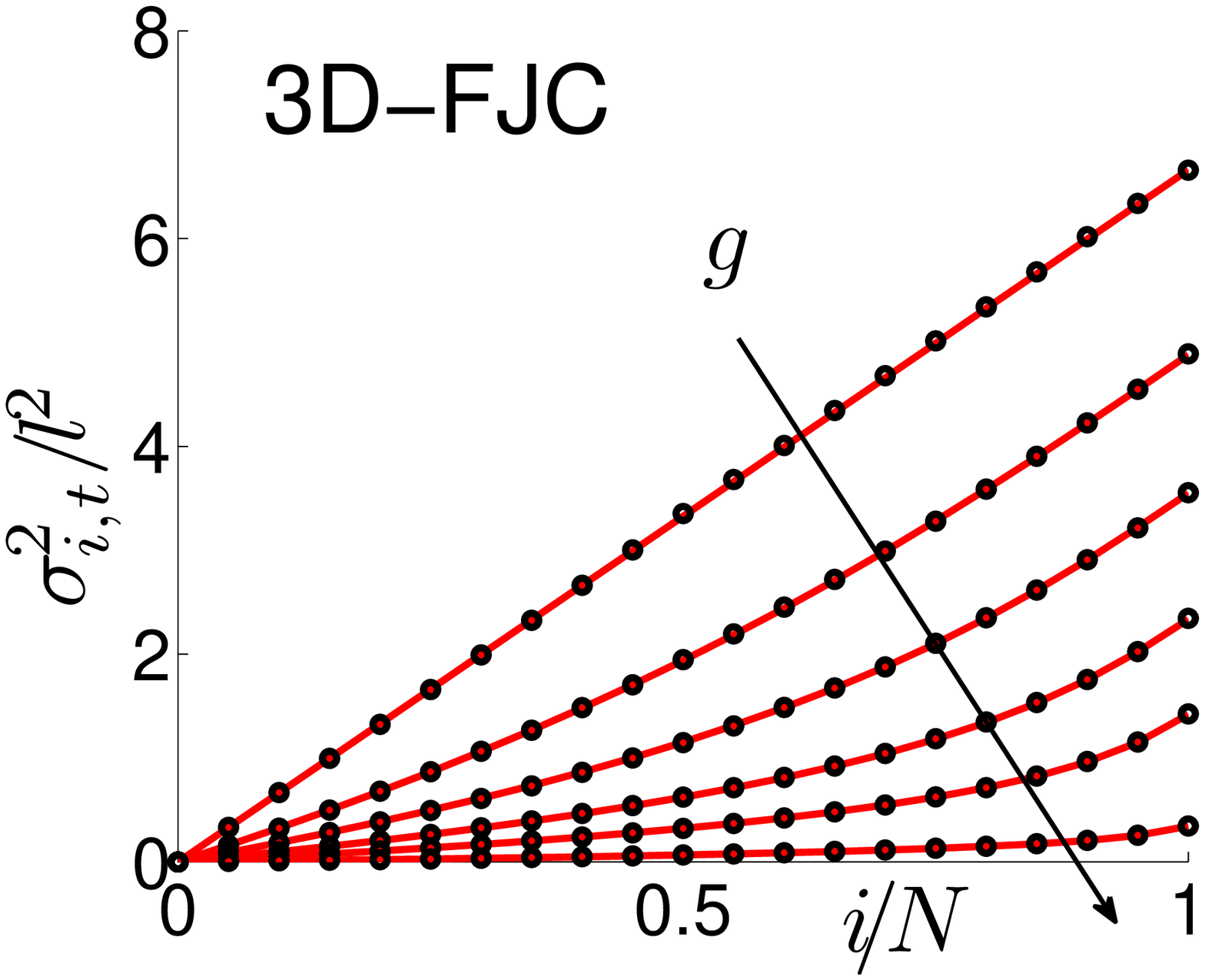}}
 \caption{(color online) Longitudinal (top panel) and transversal (bottom panel) component of the variance of positions for the 3D FJC case. The red solid lines correspond to the analytical result Eq.(\ref{variance}), MC results are superimposed in black circles. Each curve corresponds to different values of the external field amplitude defined by $gl/(k_B T)=0.1, 0.25, 0.5, 1, 2, 10$ for a fixed chain length $N=20$.
 }
 \label{variances_3D_Gamp}       
 \end{figure}

We report in Fig.\ref{variances_3D_N} and Fig.\ref{variances_3D_Gamp} the longitudinal and transversal component of the variance as a function of the chain length and the field strength for the 3D case (with $f=0$). The 2D case is very similar and it has not been reported here for sake of brevity. We can observe some interesting trends: the longitudinal variance of the position is a decreasing function of the number of polymers $N$ while the transversal one is a increasing function (with a fixed amplitude of the external field $g$). Moreover, both variances are rapidly increasing along the chain, assuming the largest value in the last free monomer, which is more subject to strong fluctuations. It interesting to observe that the variance (both longitudinal and transversal components) is a linear function of the position $i$ along the chain (it linearly intensifies along the chain itself) with a simple force $f$ applied at the free end: conversely, with a uniform field $g$, the distribution of forces generates a strongly non-linear intensification of the variances moving towards the free end-terminal. So, from the point of view of the variances, the application of a field or the application of a single force generates completely different responses. In Fig.\ref{variances_3D_Gamp} we can  also observe that the variances are decreasing functions of the strength of the field (both for the longitudinal and transversal components); in fact, the intensity of the fields tends to reduce the fluctuations of the chain, increasing, at the same time, the tension within the bonds.

\section{Worm-like chain model under external field}

In previous Sections we treated systems described by the FJC model, characterized by the complete flexibility of the chain and, therefore, by the absence of any bending contribution to the total energy. Nevertheless, in many polymer chains, especially of biological origin, the specific flexibility (described by the so-called persistence length\cite{kamien2}) has a relevant role in several bio-mechanical processes. 
In order to take into consideration these important features, with relevant applications to bio-molecules and bio-structures, in this Section we introduce the semi-flexible polymer chain characterized by a given bending energy added to the previous Hamiltonian
\begin{eqnarray}
\label{hwlc}
H &=& \sum_{i=1}^{N} \frac{\vec p_i \cdot \vec p_i}{2m} +\frac{1}{2}k\sum_{K=1}^{N}\left( \Vert \vec{r}_K-\vec{r}_{K-1}\Vert-l\right)^{2} \\
\nonumber
&&+\frac{1}{2}\kappa\sum_{i=1}^{N-1}\left(\vec{t}_{i+1}-\vec{t}_{i} \right)^{2} - \sum_{K=1}^{N} \vec g_K \cdot \vec r_K - \vec f \cdot \vec r_N
\end{eqnarray}
where $\kappa$ is the bending stiffness, $ k $ is the stretching modulus and $ \vec{t}_{i}=(\vec{r}_{i+1}-\vec{r}_{i})/\Vert \vec{r}_{i+1}-\vec{r}_{i}\Vert $ is the unit vector collinear with the $i$-th bond (see Ref.\onlinecite{manca} for details). In particular we take into consideration the classical WLC model, describing an inextensible semi-flexible chain: it means that the spring constant  $ k $ is set to a very large value (ideally $k\rightarrow\infty$) so that the bond lengths remain fixed at the value $l$. It is well known that it is not possible to calculate the partition functions in closed form for the WLC polymers. Nevertheless, some standard approximations exist for such cases leading to simple expressions for the force-extension curves when a single force $f$ is applied to one end of the chain. In the following, starting from these results, we search for the force-extension curves when the polymers is stretched through a constant field $g$.

We start with the result for the 2D-WLC with an applied force $f$: the approximated force extension curve is given by \cite{woo}
\begin{eqnarray}
\label{markosiggia2D}
\frac{fl}{k_B T}=\frac{l}{L_{p}}\left[ \frac{1}{16(1-\zeta)^{2}}-\frac{1}{16}+\frac{7}{8}\zeta\right]
\end{eqnarray}
where $\zeta=r/(lN)$ is the dimensionless elongation and $L_p=l\kappa/(k_BT)$ is the persistence length . We suppose that such a constitutive equation is invertible through the function $\mathcal{F}$, leading to the expression $\zeta=r/(lN)=\mathcal{F}({fl}/({k_B T}))$. When $\vec{f}=0$ and $\vec g_J=\vec g$ for any $J$ we search for the 2D scalar relation between
$r$ and $g=\vert \vec{g}\vert$. As discussed in a previous section (see Eqs.(\ref{fjc2Dg}) and (\ref{fjc3Dg})), we can write
\begin{eqnarray}
 \nonumber
 \frac{r}{lN} &=& \frac{1}{N} \sum_{k=1}^{N} \mathcal{F}\left( \frac{lg}{k_BT}(N-k+1) \right) \\
  \nonumber
 & \simeq & \frac{1}{N} \int_{k=0}^{N} \mathcal{F}\left( \frac{lg}{k_BT}(N-x+1) \right) dx\\
 &= & \frac{1}{N}\frac{1}{\frac{lg}{k_BT} } \int_{\frac{lg}{k_BT}}^{\frac{lg}{k_BT}(N+1)} \mathcal{F}\left( y \right)dy
\end{eqnarray}
where we have defined the change of variable $y=\frac{lg}{k_BT}(N-x+1)$. We adopt now a second change of variable through the relation $z=\mathcal{F}(y)$ or $y=\mathcal{F}^{-1}(z)$; it leads to
\begin{eqnarray}
 \nonumber
 \frac{r}{lN} &=& \frac{1}{N}\frac{1}{\frac{lg}{k_BT} } \int_{\mathcal{F}\left(\frac{lg}{k_BT}\right)}^{\mathcal{F}\left(\frac{lg}{k_BT}(N+1)\right)} z\frac{\mathcal{F}^{-1}\left( z \right)}{dz}dz\\
 \label{wlc2Dg}
 &=&\frac{1}{N}\frac{1}{\frac{lg}{k_BT} }\frac{l}{L_{p}}\\
 \nonumber
 &\times &\left[\frac{7}{16}z^{2}-\frac{1}{8(1-z)}+\frac{1}{16(1-z)^{2}} \right]_{\mathcal{F}\left(\frac{lg}{k_BT}\right)}^{\mathcal{F}\left(\frac{lg}{k_BT}(N+1)\right)} 
\end{eqnarray}
where we used the notation $[h(z)]_a^b=h(b)-h(a)$.
This result represents (although in implicit form) the approximated force-extension curve for the 2D-WLC under external fields. To evaluate Eq.(\ref{wlc2Dg}) we need to know the inverse function $\mathcal{F}(\cdot)$, a task that can be performed numerically.

\begin{figure}[ht]
 \resizebox{0.85\columnwidth}{!}{\includegraphics{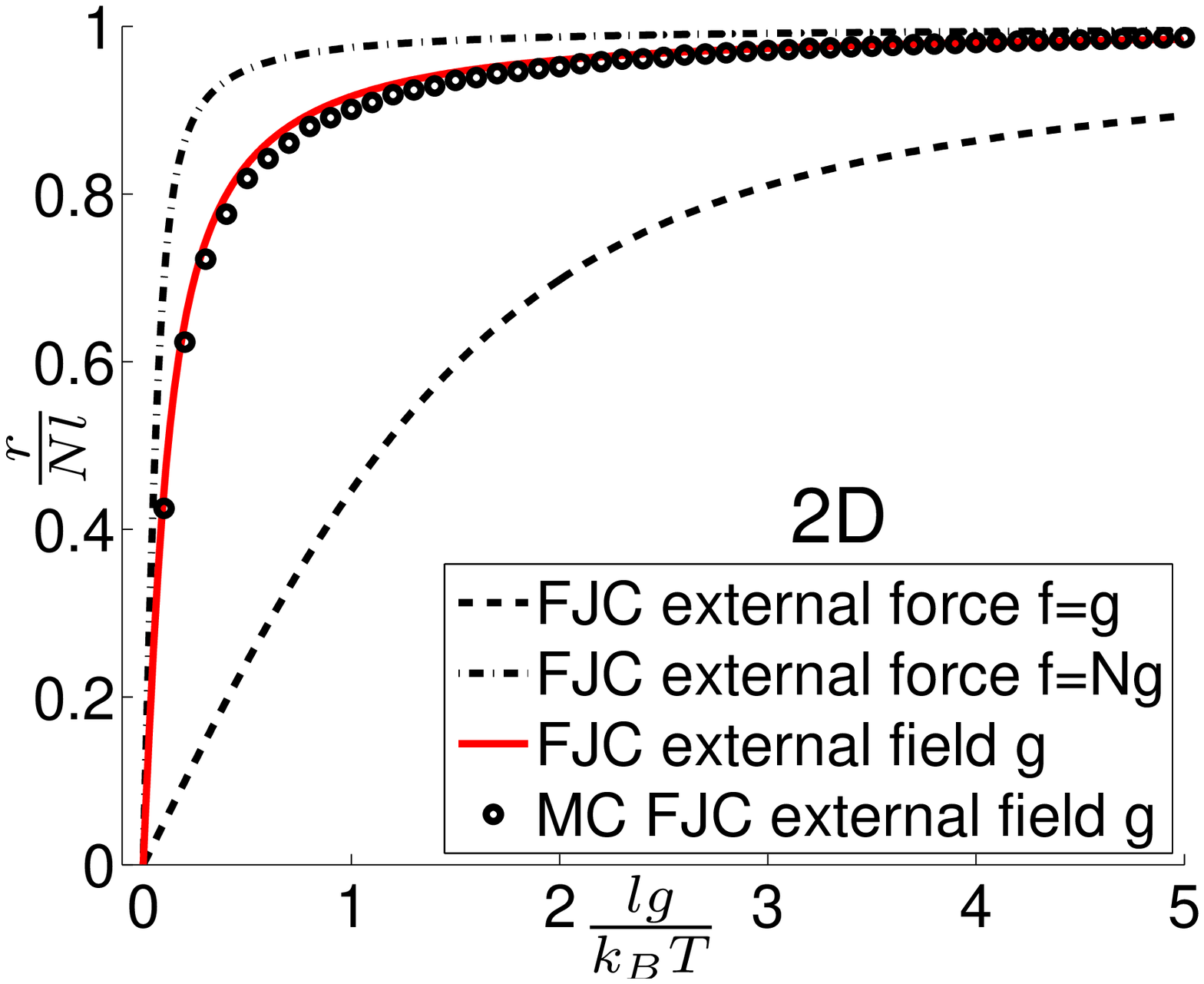}}\\
 \resizebox{0.85\columnwidth}{!}{\includegraphics{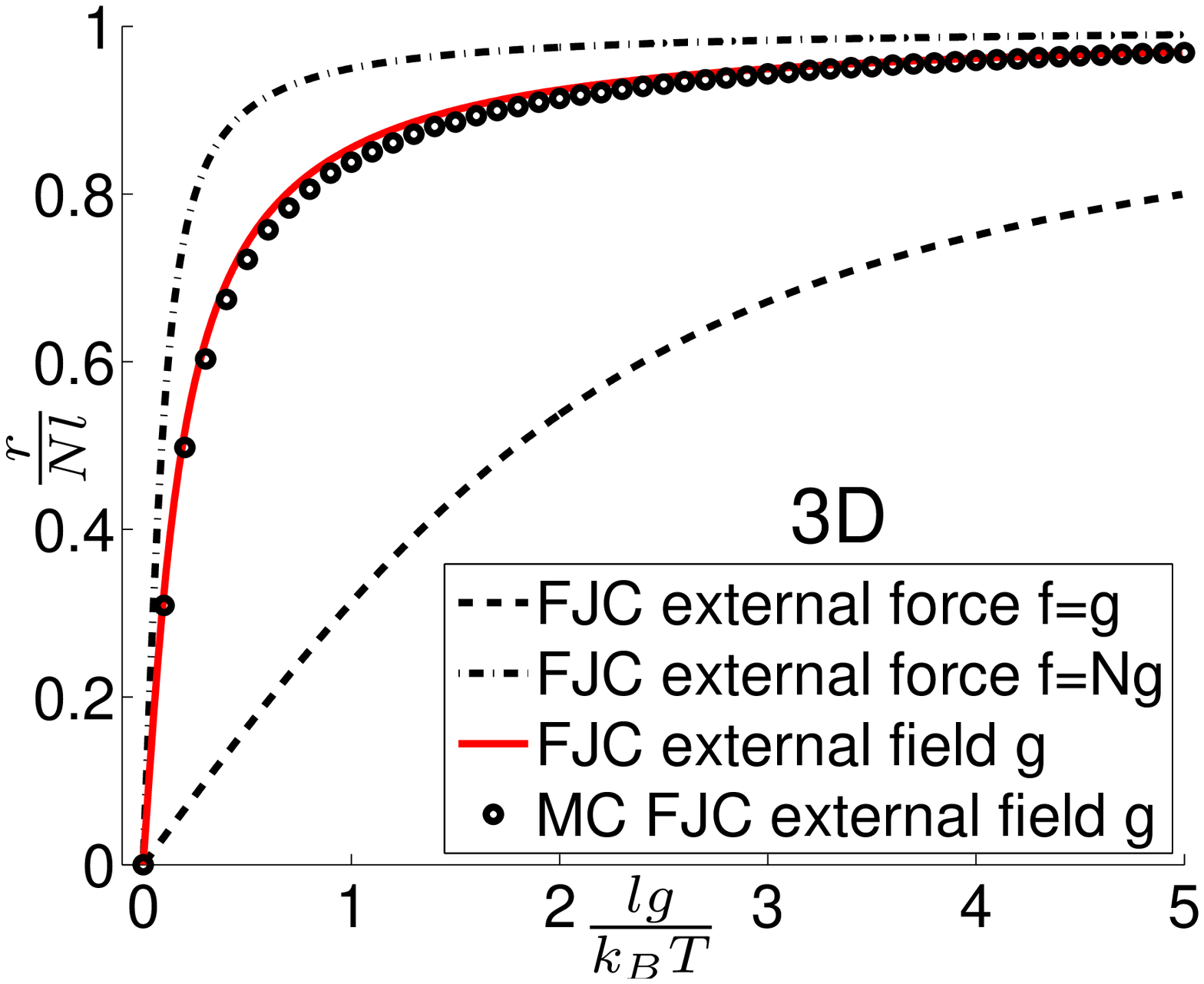}}
 \caption{(color online) Force-extension curves of a FJC polymer in an external field (or external force) with N=20. The red line corresponds to the approximated expressions given in Eqs.(\ref{fjc2Dg}) and Eqs.(\ref{fjc3Dg}) while the black circles have been obtained through MC simulations. The 2D (Eq.(\ref{fjc2Df})) and 3D (Eq.(\ref{fjc3Df})) FJC expressions (without an external field) are plotted for comparison with $f=g$ and $f=Ng$.
 }
 \label{forcextension_FJC}       
 \end{figure}
\begin{figure}[ht]
 \resizebox{0.85\columnwidth}{!}{\includegraphics{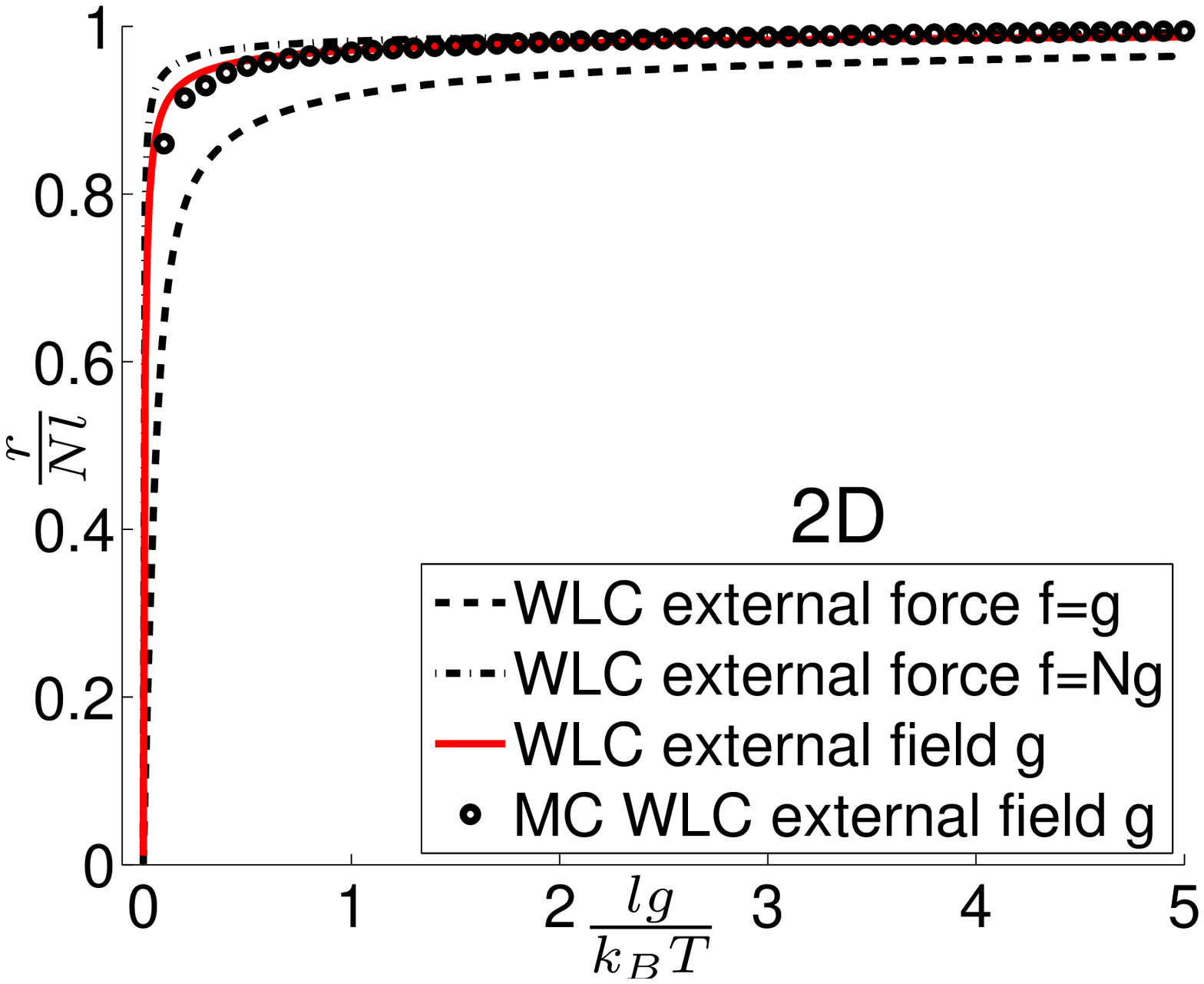}}\\
 \resizebox{0.85\columnwidth}{!}{\includegraphics{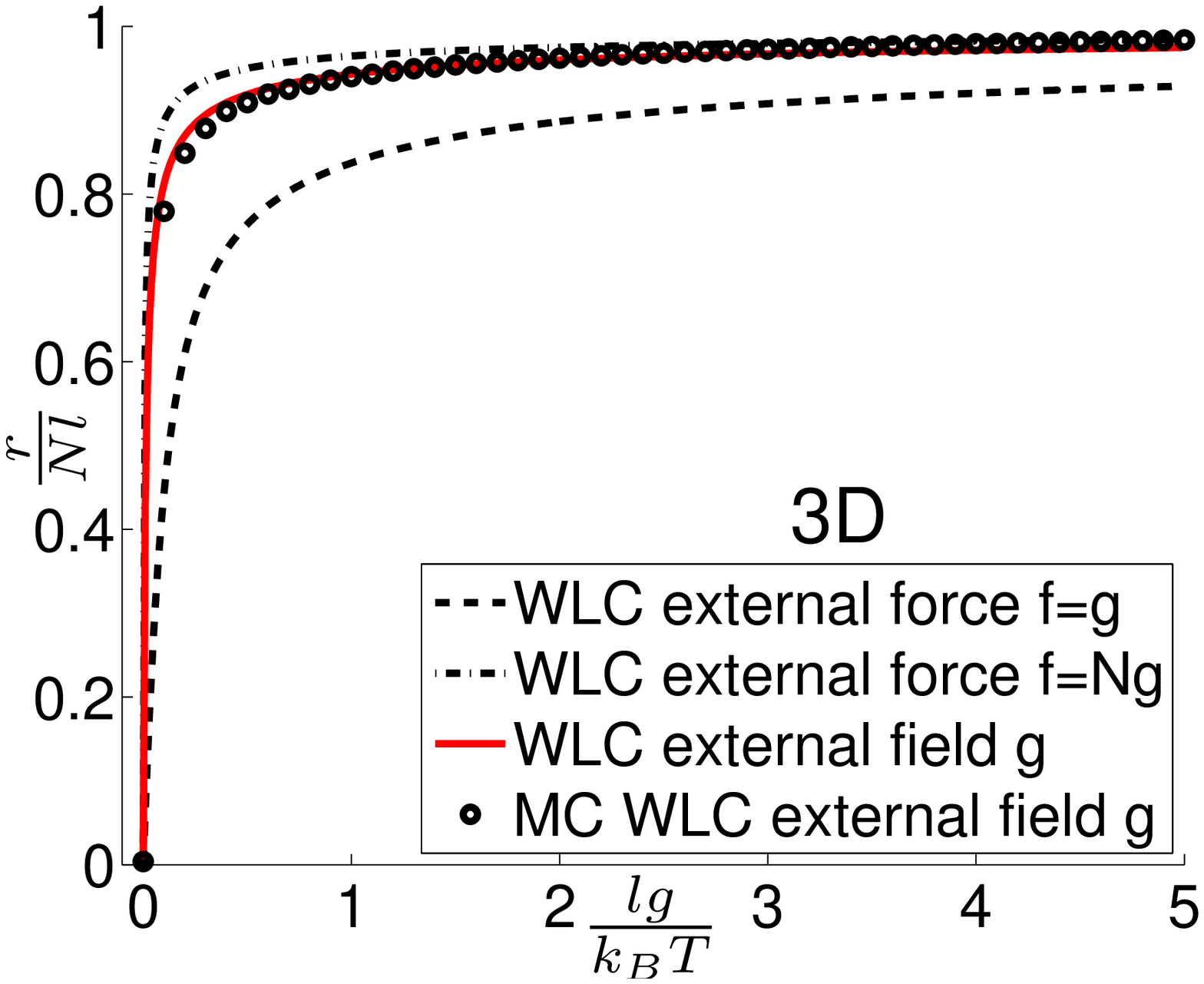}}
 \caption{(color online) Force-extension curves of a WLC polymer in an external field (or external force) with N=20. The red line corresponds to the approximated expressions given in Eqs.(\ref{wlc2Dg}) and Eqs.(\ref{wlc3Dg}) while the black circles have been obtained through MC simulations. The 2D (Eq.(\ref{markosiggia2D})) and 3D (Eq.(\ref{markosiggia3D})) WLC expressions (without an external field) are plotted for comparison with $f=g$ and $f=Ng$. The value of the bending spring constant is $\kappa = 0.4 \cdot 10^{-19} $ Nm $\simeq\,10 k_BT$ at $T=293$K. }
 \label{forcextension_WLC}       
 \end{figure}

Similarly, we may consider the standard 3D-WLC model with an applied force $f$; the classical Marko-Siggia result\cite{marko} is
\begin{eqnarray}
\label{markosiggia3D}
\frac{fl}{k_B T}=\frac{l}{L_{p}}\left[ \frac{1}{4(1-\zeta)^{2}}-\frac{1}{4}+\zeta\right]
\end{eqnarray}
where, as before,  $\zeta=r/(lN)$ is the dimensionless elongation and $L_p=l\kappa/(k_BT)$ is the persistence length. We suppose again that such constitutive equation is invertible through the function $\mathcal{G}$, leading to the expression $\zeta=r/(lN)=\mathcal{G}({fl}/({k_B T}))$. When $\vec{f}=0$ and $\vec g_J=\vec g$ for any $J$ we search for the 3D scalar relation between
$r$ and $g=\vert \vec{g}\vert$. By repeating the previous procedure, we can write
\begin{eqnarray}
 \nonumber
 \frac{r}{lN} &=& \frac{1}{N}\frac{1}{\frac{lg}{k_BT} } \int_{\mathcal{G}\left(\frac{lg}{k_BT}\right)}^{\mathcal{G}\left(\frac{lg}{k_BT}(N+1)\right)} z\frac{\mathcal{G}^{-1}\left( z \right)}{dz}dz\\
  \label{wlc3Dg}
 &=&\frac{1}{N}\frac{1}{\frac{lg}{k_BT} }\frac{l}{L_{p}}\\
 \nonumber
 &\times &\left[\frac{1}{2}z^{2}-\frac{1}{2(1-z)}+\frac{1}{4(1-z)^{2}} \right]_{\mathcal{G}\left(\frac{lg}{k_BT}\right)}^{\mathcal{G}\left(\frac{lg}{k_BT}(N+1)\right)} 
\end{eqnarray}
which represents the implicit form of the approximated force-extension curve for the 3D-WLC under external fields.

\begin{table}
\caption{Asymptotic forms of the force-extension curves for all cases described in the paper: FJC and WLC models in  2D and 3D geometry with force applied $f$ or field applied $g$.\label{asym}} 
\begin{ruledtabular}
\begin{tabular}{l c  c}
 & Asymptotic form & Asymptotic form \\
$\underbrace{\mbox{Polymer chain}}_{Equation}$ & of $ \frac{r}{lN}$ for $f,g\rightarrow 0$ & of $ \frac{r}{lN}$ for $f,g\rightarrow \infty$  \\
 & $\left(x=\frac{lf}{k_BT}\mbox{ or }\frac{lg}{k_BT} \right) $ & $\left(x=\frac{lf}{k_BT}\mbox{ or }\frac{lg}{k_BT} \right) $ \\
\hline
\\
$\underbrace{\mbox{FJC (2D) }f}_{Eq.(\ref{fjc2Df})}$     & $\dfrac{1}{2}x$      & $1-\dfrac{1}{2x}$  \\ \\
$\underbrace{\mbox{FJC (3D) }f}_{Eq.(\ref{fjc3Df})}$     &  $\dfrac{1}{3}x$      &  $1-\dfrac{1}{x}$  \\ \\
$\underbrace{\mbox{FJC (2D) }g}_{Eq.(\ref{fjc2Dg})}$     &  $\dfrac{1}{2}\left(1+\dfrac{N}{2} \right) x$      & $1-\dfrac{\log(N+1)}{2N}\dfrac{1}{x}$   \\ \\
$\underbrace{\mbox{FJC (3D) }g}_{Eq.(\ref{fjc3Dg})}$ &  $\dfrac{1}{3}\left(1+\dfrac{N}{2} \right) x$    & $1-\dfrac{\log(N+1)}{N}\dfrac{1}{x}$    \\ \\
$\underbrace{\mbox{WLC (2D) }f}_{Eq.(\ref{markosiggia2D})}$ &  $\dfrac{L_p}{l}x$ &  $1-\dfrac{1}{4}\dfrac{1}{\sqrt{\dfrac{L_p}{l}x}}$  \\ \\
$\underbrace{\mbox{WLC (3D) }f}_{Eq.(\ref{markosiggia3D})}$& $\dfrac{2}{3}\dfrac{L_p}{l}x$   &  $1-\dfrac{1}{2}\dfrac{1}{\sqrt{\dfrac{L_p}{l}x}}$ \\ \\
$\underbrace{\mbox{WLC (2D) }g}_{Eq.(\ref{wlc2Dg})}$& $\dfrac{L_p}{l}\left(1+\dfrac{N}{2} \right)x$    &  $1-\dfrac{1}{\sqrt{\dfrac{L_p}{l}x}}\dfrac{\sqrt{N+1}-1}{2N}$   \\ \\
$\underbrace{\mbox{WLC (3D) }g}_{Eq.(\ref{wlc3Dg})}$& $\dfrac{2}{3}\dfrac{L_p}{l}\left(1+\dfrac{N}{2} \right)x$    &  $1-\dfrac{1}{\sqrt{\dfrac{L_p}{l}x}}\dfrac{\sqrt{N+1}-1}{N}$ \\
\end{tabular}
\end{ruledtabular}
\end{table}

It is interesting to compare the very different force-extension curves for a single molecule in the two cases of a uniform (only $f$ applied) and non-uniform (only $g$ applied) stretch. In particular, taking advantage of our approximated formulas, we can analyse the case of a FJC and a WLC polymer. The 2D and 3D FJC results are plotted in Fig.\ref{forcextension_FJC}; on the other hand, the 2D and 3D WLC curves have been shown in Fig.\ref{forcextension_WLC}. 
For the WLC case we assumed $\kappa=10 k_B T$ for the bending modulus at $T=293$K. This value is comparable to that of polymer chains of biological interest (e.g., for DNA $\kappa=15 k_B T$).\cite{marko}
In any case three curves have been reported for drawing all the possible comparisons: the response under the field $g$, the response under the force $f=g$ and, finally, the response to an external force $f=Ng$. 
Interesting enough we note that the curve corresponding to the field $g$ is always comprised between the cases with only the force $f=g$ and $f=Ng$. The response with the field $g$ is clearly larger than that with the single force $f=g$ since the field corresponds to a distribution of $N$ forces (of intensity $f$) applied to all monomers; therefore, the total force applied is larger, generating a more intense effect. 
However, the case with a single force $f=Ng$ shows a response larger than that of the field $g$. 
In this case the total force applied in the two cases is the same but the single force $Nf$ is applied entirely to the last terminal monomer, generating an overall stronger effect compared to the same force evenly distributed on the monomers. In fact, a force generates a stronger effect if it is placed in the region near the free polymer end (its effect is redistributed also to all preceding bonds). 
The curves in   Fig.\ref{forcextension_FJC} and Fig.\ref{forcextension_WLC} have been obtained  with the theoretical formulations presented in this Section and confirmed by a series of MC simulations. In all case we obtained a quite perfect agreement between the two formulations. 
The knowledge of the closed-form expressions allowed us to analytically analyze the behavior of the chains for very low and very high applied forces (or fields). The results are shown in Table \ref{asym}: interestingly, we note that the extension is always a linear function of the small applied perturbation. 
Nevertheless, the corresponding constant of proportionality depends on $N$ only when a field is applied to the chain; conversely, it is independent of $N$ with a single force applied at one end. On the other hand, with a large perturbation applied to the molecule, we observe a $1/x$ behavior for the FJC models and a $1/\sqrt{x}$ behavior for the WLC models. 
To conclude we also remark that the order of the curves observed in  Fig.\ref{forcextension_FJC} and Fig.\ref{forcextension_WLC} is confirmed also in the low and high force (or field) regime by the following inequalities: $1<1+N/2<N$ (low force regime) and $1<\log(N+1)<N$ (high force regime) for the FJC model and $1<1+N/2<N$ (low force regime) and $\sqrt{N}<2(\sqrt{N+1}-1)<N$ (high force regime) for the WLC model (always for $N\geq 2$).

\section{Action of a pulling force not aligned with the external field}

 \begin{figure}
  \hspace{1cm}\resizebox{0.7\columnwidth}{!}{\includegraphics{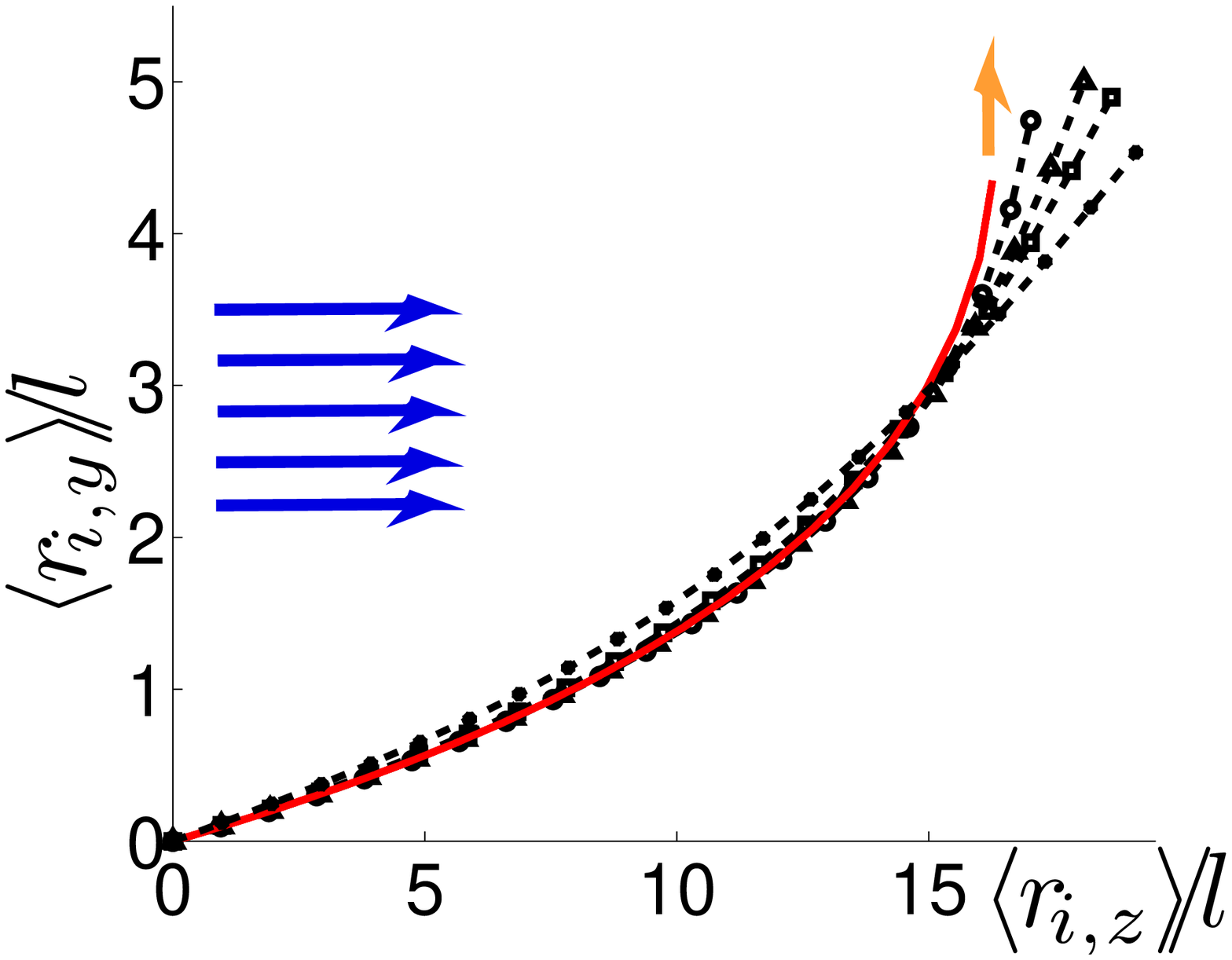}}
   \resizebox{0.7\columnwidth}{!}{\includegraphics{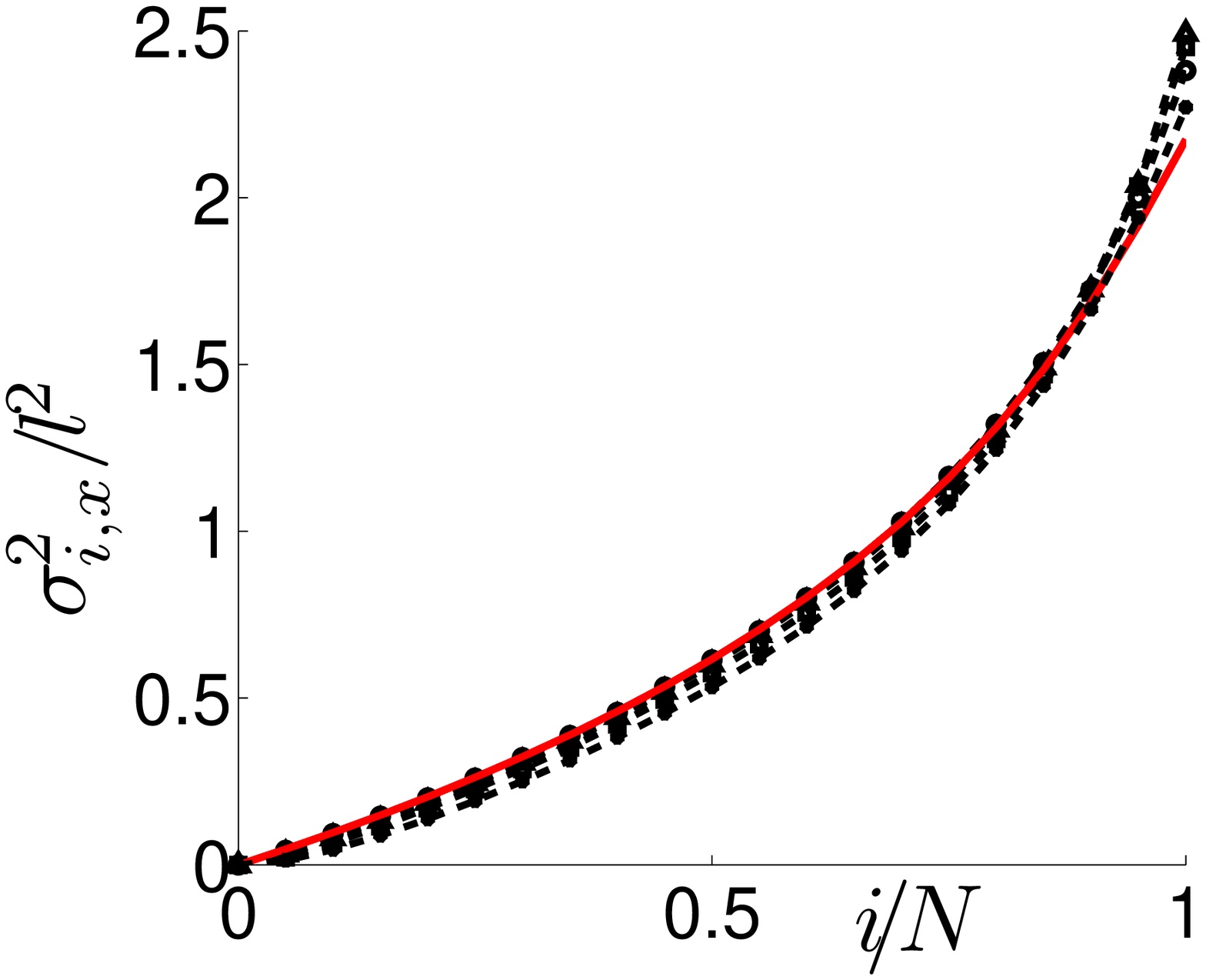}}
  \resizebox{0.7\columnwidth}{!}{\includegraphics{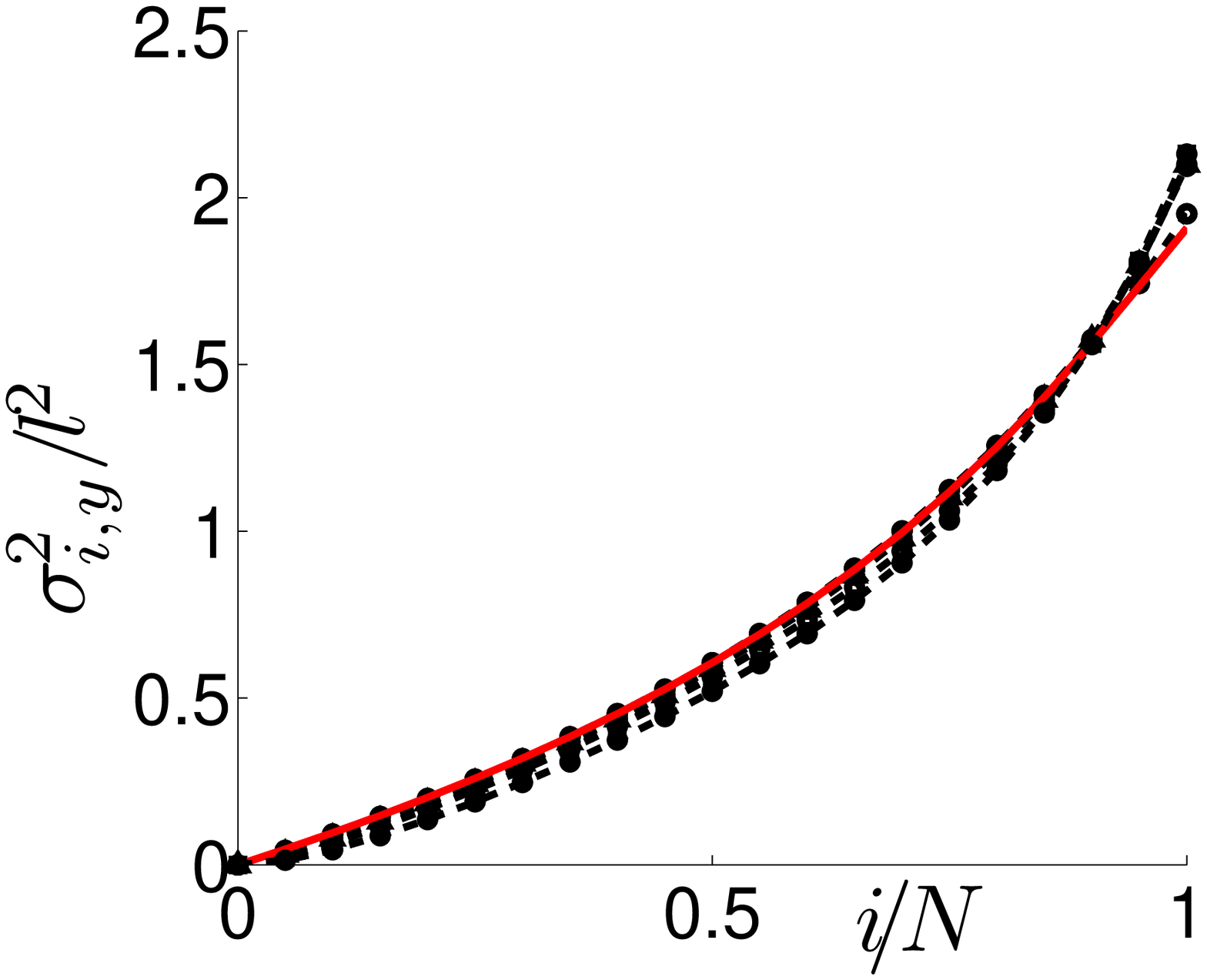}}
  \resizebox{0.7\columnwidth}{!}{\includegraphics{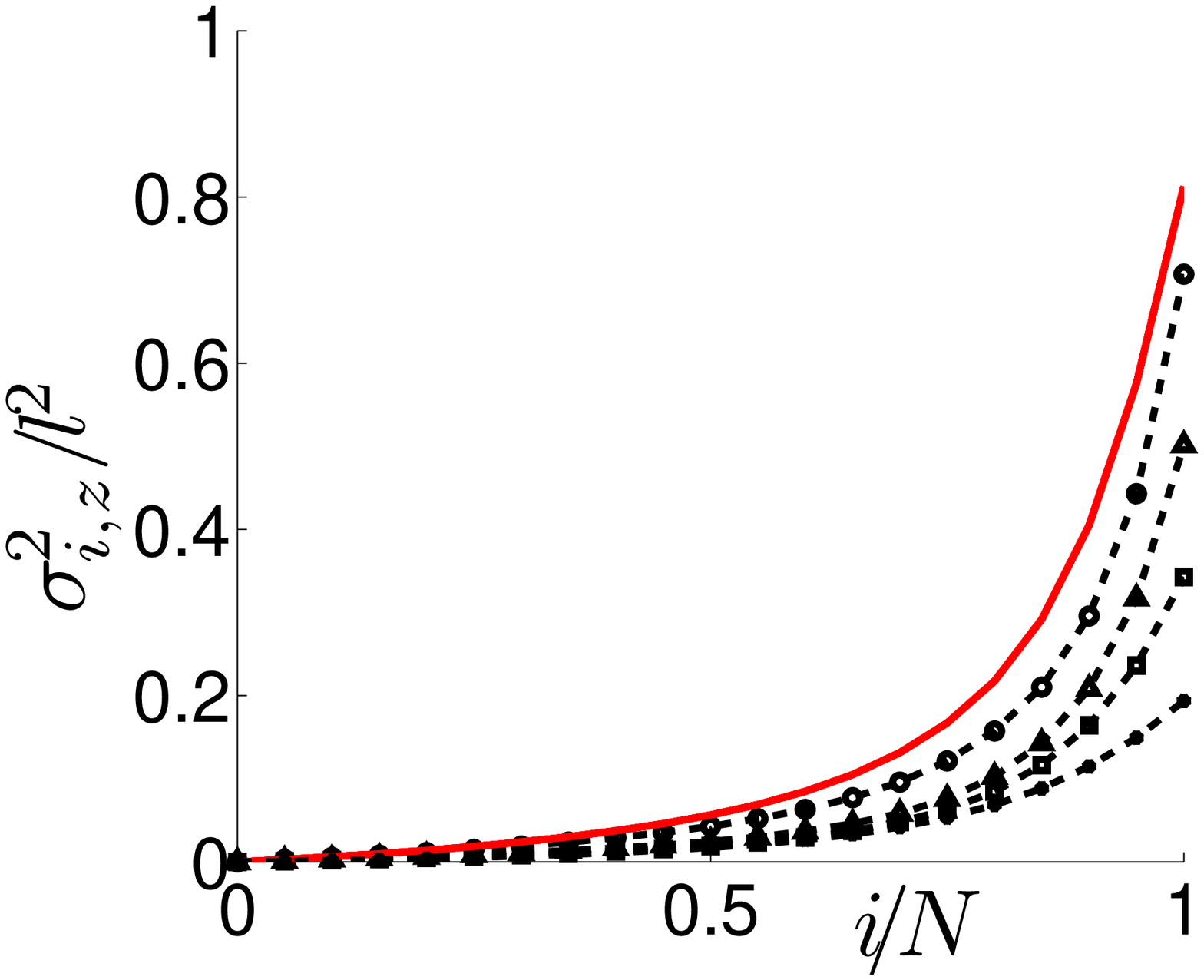}}
  \caption{(color online) Action of a pulling force $f$ (along the $y$-axis) perpendicular to the applied field $g$ (along the $z$-axis). We adopted different values of the bending spring constant: $\kappa = 0.08, 0.6, 2, 8 \cdot 10^{-19} $ Nm. The chain length is fixed $(N=20)$, the external field amplitude is $g=4$ pN and the force applied to the last monomer of the chain corresponds to $f=8$ pN. The red solid lines correspond to the analytical results for the FJC case (see Eqs.(\ref{3Dcampo}) and (\ref{variance})). Black circles correspond to the MC simulations with the different bending spring constants. In the top panel we reported the average positions, while in the others the three variances of the $x$, $y$ and $z$ components.}
  \label{tang}       
  \end{figure}

\begin{figure}
 \resizebox{0.7\columnwidth}{!}{\includegraphics{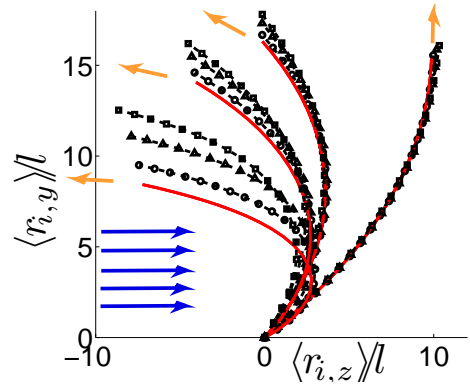}}
 \caption{(color online) Average positions of the chain for different angles between the external traction force $f$ and the direction of the applied field $g$. We adopted $N=20$, $g=4$ pN and $f=60$ pN.  The red solid lines correspond to the FJC analytical result, Eq.(\ref{3Dcampo}). The symbols represent the MC results for the WLC model with $\kappa = 0.08, 0.6, 2 \cdot 10^{-19} $ Nm (circles, triangles and squares, respectively). For both FJC and WLC models we used different values of the angle between the applied field and the traction force  $\theta= \pi/2, 3\pi/4, 5\pi/6, 15\pi/16$ from the right left. }
 \label{pos_3D_bending}       
 \end{figure}

 \begin{figure*}
  \resizebox{0.67\columnwidth}{!}{\includegraphics{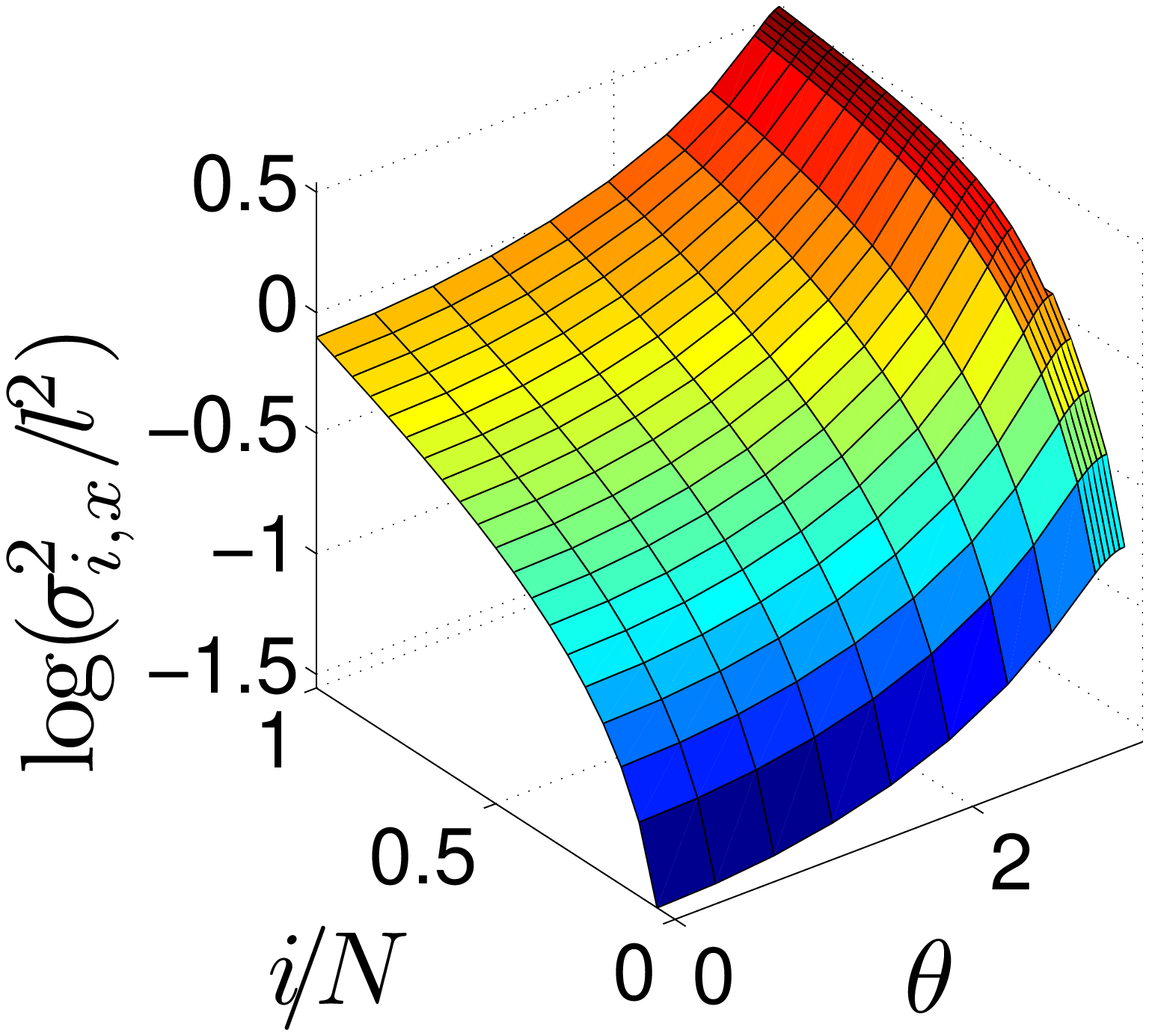}}
  \resizebox{0.67\columnwidth}{!}{\includegraphics{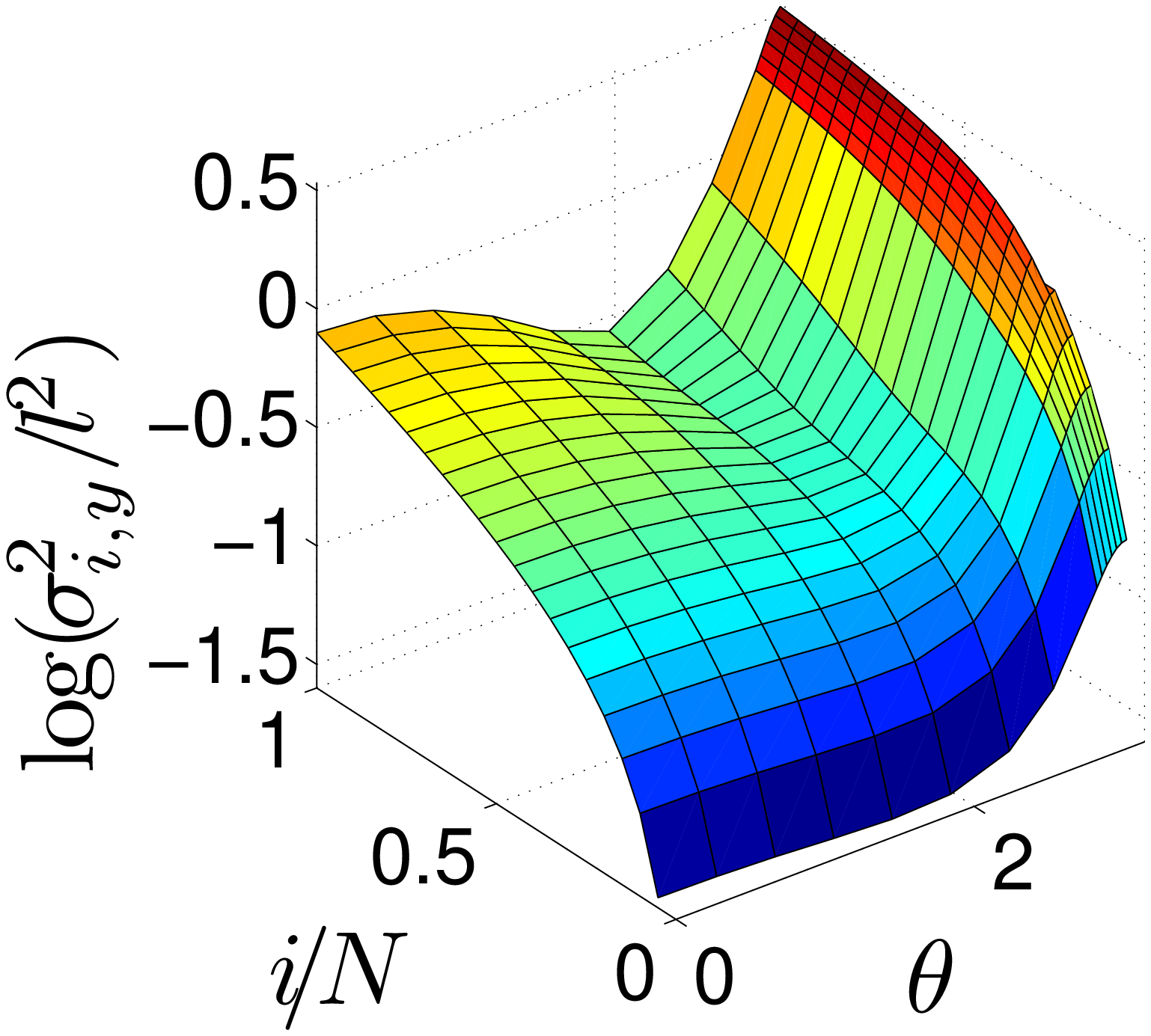}}
  \resizebox{0.67\columnwidth}{!}{\includegraphics{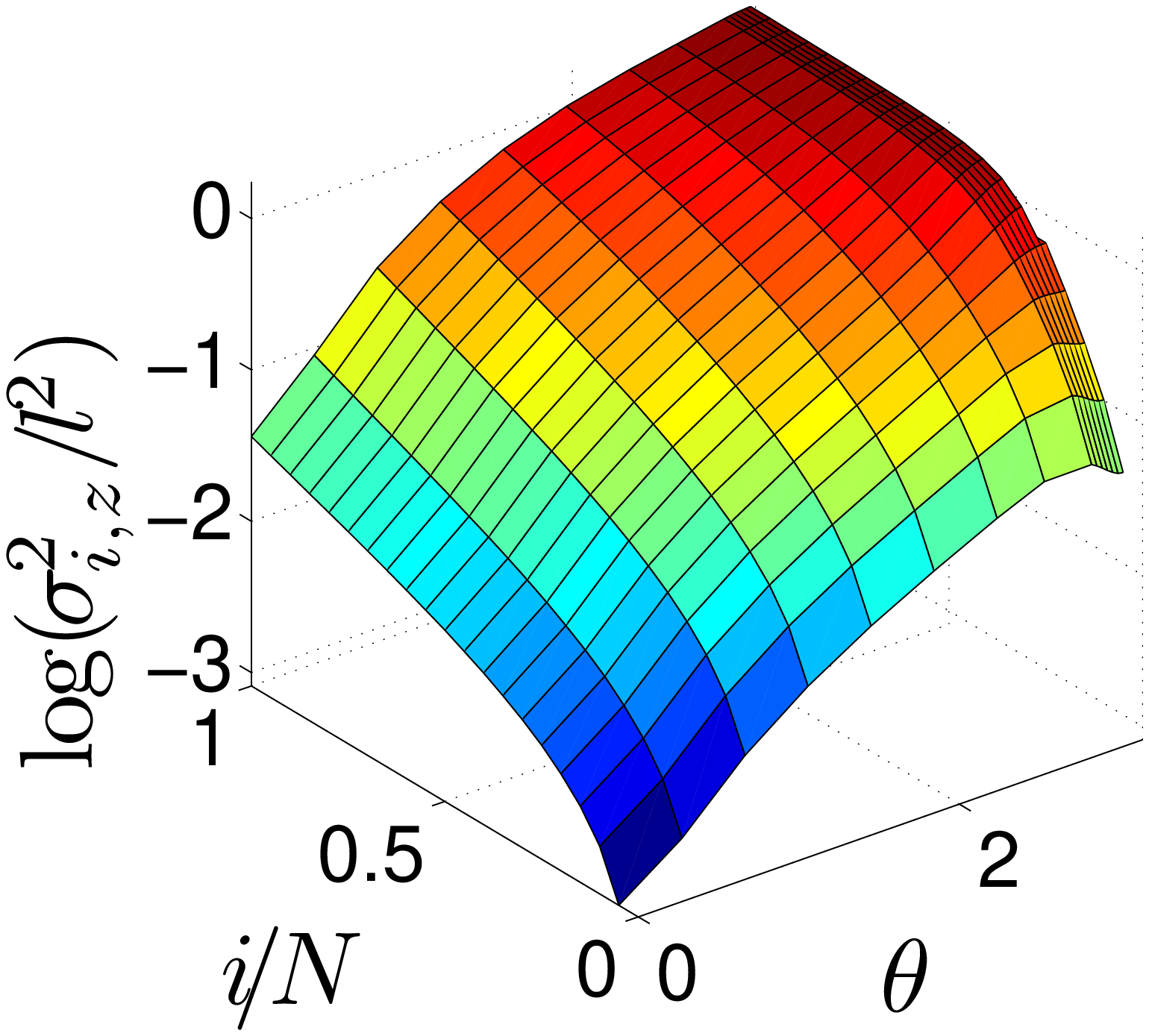}}
  \caption{(color online) Monomer variances versus the position along the chain ($i$) and the angle between force and field ($0<\theta<\pi$) for the FJC model. As before we used $N=20$, $g=4$ pN and $f=60$ pN.  }
  \label{var-fjc}       
  \end{figure*}

\begin{figure*}
 \resizebox{0.67\columnwidth}{!}{\includegraphics{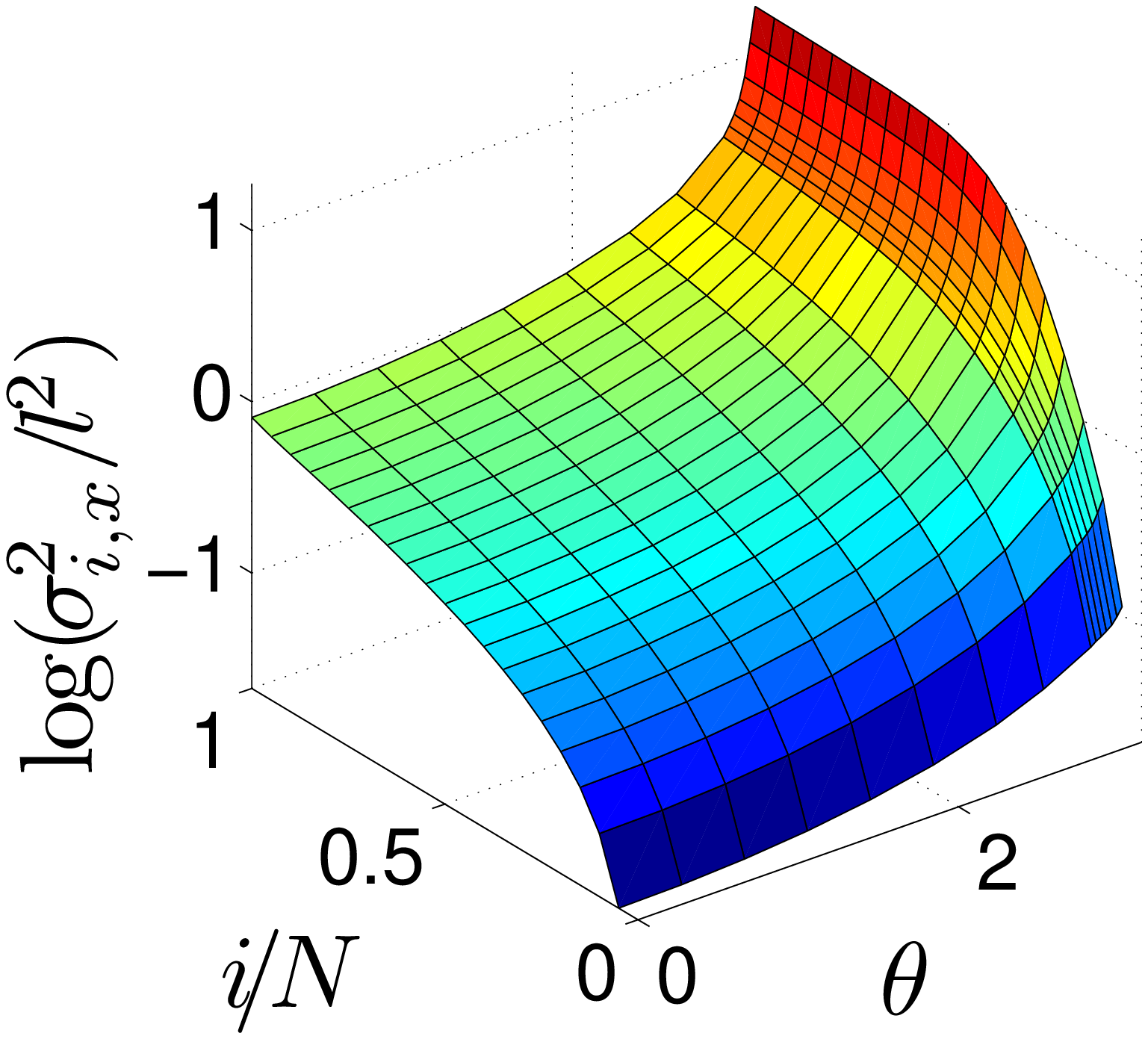}}
 \resizebox{0.67\columnwidth}{!}{\includegraphics{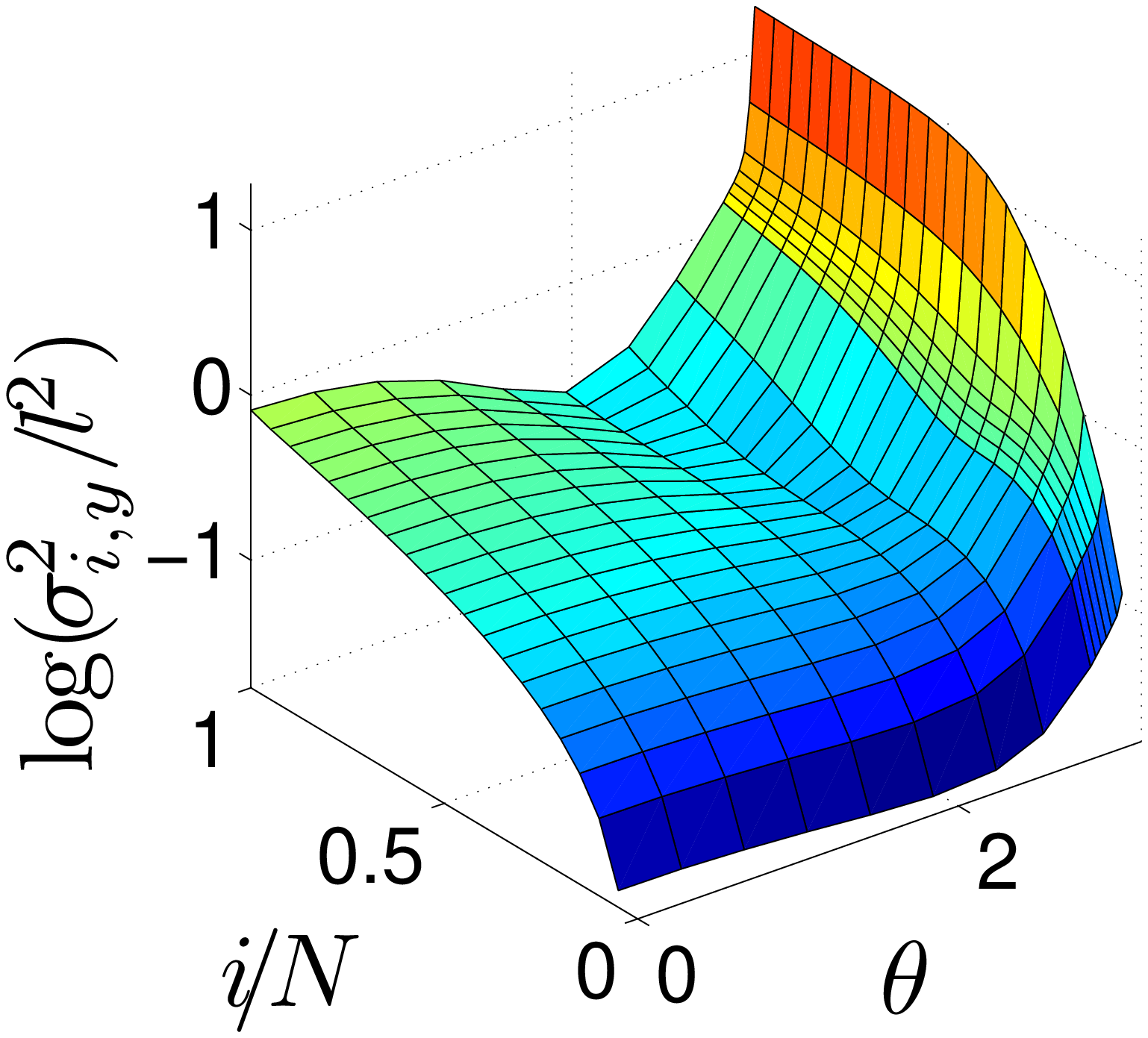}}
 \resizebox{0.67\columnwidth}{!}{\includegraphics{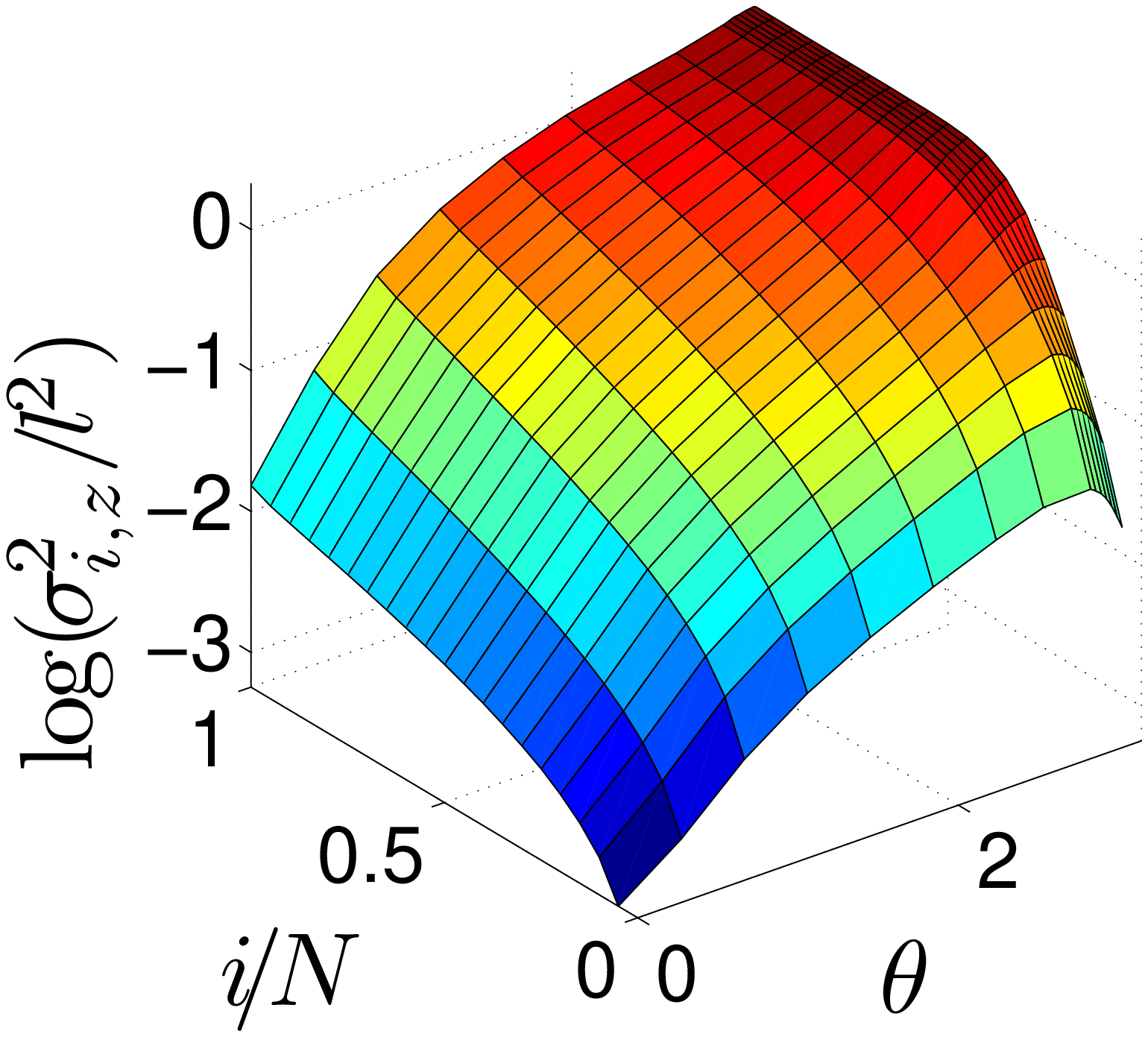}}
 \caption{(color online) Monomer variances versus the position along the chain ($i$) and the angle between force and field ($0<\theta<\pi$) for the WLC model. As before we used $N=20$, $g=4$ pN and $f=60$ pN. We also adopted a bending stiffness $\kappa = 0.6 \cdot 10^{-19} $ Nm.  }
 \label{var-wlc}       
 \end{figure*}

In previous Sections we considered the polymer chain immersed in an external field with an external force equal to zero at its end. However, since we developed a form of the partition function also taking into account an external force applied at the end of the chain (at least for the FJC model), we can directly study the important case with a non zero force superimposed to an external field, in general having different orientation.  To do this, we keep fixed the origin of the chain and apply a constant force at the end of the polymer with different angles with respect to the direction of the applied field. We will analyse such a problem for both the FJC and WLC cases. 

To begin, we consider a pulling force perpendicular to the direction of the applied field, respectively the $y$ and $z$ axis of our reference frame. For increasing values of the bending spring constant $\kappa$ going from nearly zero (FJC model) to $8 \cdot 10^{-19} $Nm (WLC model, including the bending constant of the DNA given by $\kappa=0.6 \cdot 10^{-19} $ Nm $\simeq\,15 k_B T$). In Fig.\ref{tang} we reported the results for the average monomers positions and their variances.  The red solid lines correspond to the analytical results for the FJC case, while the black symbols correspond to the MC simulations. It is interesting to observe the effect of the persistence length (or, equivalently of the bending stiffness): in fact, in the top panel of Fig.\ref{tang} we note that the chains with an higher bending spring constant tend to remain more straight under the same applied load. At the same time, in the fourth panel of Fig.\ref{tang} we observe a decreasing variance along the $z$-axis (direction of the applied field) with an increasing bending spring constant; this fact can be easily interpreted observing that an higher rigidity of the chain reduces the statistical fluctuations in the direction of the applied field. The situation is more complicated for the variances along the $x$ and $y$ directions: in fact, along the chain, there are some monomers with variances larger than the corresponding FJC case and others with smaller values.

In Fig.\ref{pos_3D_bending} the average positions of the monomers for different directions of the external force are reported. The figure shows how the average monomer positions depend on the bending rigidity $\kappa$ and on the external force angle $\theta$. 
As before we can observe that the persistence length of the chain tends to maintain a low curvature in the shape of the chain. This phenomenon is more evident with an increasing angle between the force and the field. 
In fact, in Fig.\ref{pos_3D_bending}, the deviation between the FJC results and the WLC ones is higher for the angles approaching $\pi$, where the force and the field are applied in opposite directions. 

In Figs.\ref{var-fjc} and \ref{var-wlc} the three components of the variance are reported versus the position of the monomer along the chain and the angle between the field and the force directions, for the FJC and WLC case, respectively. We can extract some general rules about this very complex scenario: 
as for the variance along the $x$ direction we observe it to be an increasing function both of the position $i$ along the chain an of the angle $\theta$ between $f$ and $g$. Both behaviors can be interpreted with the concept of persistence length, as discussed above. 
Conversely, the description of the variance along the $y$ direction is more complicated. In fact, while the increasing trend of the variance with the position $i$ along the chain is maintained, we observe a non monotonic behavior in terms of the angle $\theta$, with a minimum of the variance at about $\theta=2\pi/3$. Finally, the variance along the $z$ direction is always increasing along the chain, but it shows a maximum near $\theta=\pi$ (at least in the first part of the polymer chain).

\section{Conclusions}
In this work we investigated mechanical and conformational properties of flexible and semi-flexible polymer chains in external fields.
As for the FJC model we developed a statistical theory, based on the exact analytical determination of the partition function, which generalizes previous results to the case where an external field is applied to the system. In particular we obtained closed form expression for both the average conformation of the chain and its covariance distribution. For sake of completeness, all calculations have been performed both in two-dimensional and three-dimensional geometry. On the other hand, as for the WLC model we derived new approximate expressions describing the force-extension curve under the effect of an external field. They can be considered as the extensions of the classical Marko-Siggia relationships describing the polymer pulled by a single external force applied at the free end of the chain. 
All our analytical results, for both FJC and WLC models, have been confirmed by a series of Monte Carlo simulations, always found in very good agreement with the theory.  

The overall effects generated on the tethered polymer by the application of an external field can be summarized as follows. 
As for the average configuration of a chain, it is well known that a single pulling force generates a uniform deformation along the chain (for a homogeneous polymer with all monomers described by the same effective elastic stiffness). On the contrary, the application of an external field produces a non uniform deformation along the chain, showing a larger deformation in the portion of the chain closest to the fixed end. 
Moreover, the variances of the positions increase linearly along the chain with a single force applied to the polymer. Conversely, the polymer subjected to an external field exhibits a non-linearly increasing behavior of the variances along the chain. More specifically the variances assume the largest values nearby the last free monomers, where we can measure the highest fluctuations.

To conclude, we underline that the use of the MC method, once validated against known analytical solutions, is crucial for analysing models conditions which are beyond reach of a full analytical calculation. We take full profit of this approach for analysing the effects of the combination of an applied force at the free end together with an external field, especially when the two are not aligned. We have analysed the average configurational properties of the polymer, observing a very complex scenario concerning the behavior of the variances.

\begin{acknowledgments}
We acknowledge computational support by CASPUR (Rome, Italy) under project ``Standard HPC Grant 2011/2012''. FM acknowledges the Department of Physics of the University of Cagliari for the extended visiting grant, and the IEMN for the kind hospitality offered during part of this work. 
\end{acknowledgments}




\begin{thebibliography}{}

\bibitem{lots}
A. Ashkin, Proc. Natl Acad. Sci. \textbf{94}, 4853 (1997).

\bibitem{mts}
C. Gosse and V. Croquette, Biophys. J. \textbf{82}, 3314 (2002).

\bibitem{AFM}
D. M. Czajkowsky and Z. Shao, FEBS Lett. \textbf{430}, 51 (1998).

\bibitem{nucleic}
C. R. Calladine, H. R. Drew, B. F. Luisi, and A. A. Travers, \textit{Understanding DNA: the molecule and how it works}; Elsevier Academic Press, Amsterdam, 1992.

\bibitem{bensimon}
A. Bensimon, A. Simon, A. Chiffaudel, V. Croquette, F. Heslot, and D. Bensimon, Science \textbf{265}, 2096 (1994).

\bibitem{chan}
 E. Y. Chan, N. M. Goncalves, R. A. Haeusler, A. J. Hatch, J. W. Larson, A. M. Maletta, G. R. Yantz, E. D. Carstea, M. Fuchs, G. G. Wong, S. R. Gullans, and R. Gilmanshin, Genome Res. \textbf{14}, 1137 (2004).

\bibitem{rief1}
M. Rief, F. Oesterhelt, B. Heymann, and H. E. Gaub, Science \textbf{275}, 28 (1997).

\bibitem{rief2}
M. Rief, M. Gautel, F. Oesterhelt, J. M. Fernandez, and H. E. Gaub, Science \textbf{276}, 1109 (1997).

\bibitem{mancaII}
F. Manca, S. Giordano, P. L. Palla, F. Cleri, and L. Colombo, Two-state
theory of  single-molecule stretching, submitted (2012).

\bibitem{busta0}
C. Bustamante, S. B. Smith , J. Liphardt, and D.Smith , Curr. Op. Struct. Biol. \textbf{10},  279 (2000).

\bibitem{smith1}
S. B. Smith, L. Finzi, and C. Bustamante, Science \textbf{258}, 1122 (1992).

\bibitem{marko}
J. F. Marko and E. D. Siggia, Macromolecules \textbf{28}, 8759 (1995).

\bibitem{manca}
F. Manca, S. Giordano, P. L. Palla, R. Zucca, F. Cleri, and L. Colombo, J. Chem. Phys. \textbf{136}, 154906 (2012).

\bibitem{huguet}
J. M. Huguet, C. V. Bizarro, N. Forns, S. B. Smith, C. Bustamante, and F. Ritort, Proc. Nat. Ac. Sci. (PNAS) \textbf{107}, 15341 (2010).

\bibitem{trahan}
D. W. Trahan and P. S. Doyle, Biomicrofluidics \textbf{3}, 012803 (2009).

\bibitem{wang}
S. G. Wang and Y. G. Zhu, Biomicrofluidics \textbf{6}, 024116 (2012).

\bibitem{hsieh}
C.-C. Hsieh and T.-H. Lin, Biomicrofluidics \textbf{5}, 044106 (2011).

\bibitem{schwartz}
D. C. Schwartz, Li X., L. I. Hernandez, S. P. Ramnarain, E. J. Huff, and  Y.-K. Wang, Science \textbf{262}, 110 (1993).

\bibitem{strick1}
T. R. Strick, J. F. Allemand, D. Bensimon, A. Bensimon and  V. Croquette, Science \textbf{271}, 1835 (1996).

\bibitem{strick2}
T. R. Strick, V. Croquette, and D. Bensimon, Nature \textbf{404}, 901 (2000).


\bibitem{busta_cat}
C. Bustamante, J.C. Macosko, and G.J. Wuite, Nat. Rev. Mol. Cell. Biol. \textbf{1},130 (2000).

\bibitem{smith2}
D. E. Smith, H. P. Babcock, and S. Chu, Science \textbf{283}, 1724 (1999).

\bibitem{perkins1}
T. T. Perkins, D. E. Smith, R. G. Larson, and S. Chu, Science \textbf{268}, 83 (1995).

\bibitem{perkins2}
T. T. Perkins, S. R. Quake, D. E. Smith, and S. Chu, Science \textbf{264}, 822 (1994).

\bibitem{warner}
M. Warner, J. M. F. Gunn, and A. B. Baumgartner, J. Phys. A: Math. Gen. \textbf{19}, 2215 (1986).

\bibitem{vroege}
G. J. Vroege  and T. Odijk,  Macromolecules \textbf{21}, 2848 (1988).

\bibitem{kamien}
K. D. Kamien, P. L. Doussal, and D. R. Nelson, Phys. Rev. A \textbf{45}, 8727 (1992).

\bibitem{brochard1}
F. Brochard-Wyart,  Europhys. Lett. \textbf{23}, 105 (1993).

\bibitem{brochard2}
F. Brochard-Wyart, H. Hervet, and P. Pincus,  Europhys. Lett. \textbf{26}, 511 (1994).

\bibitem{brochard3}
F. Brochard-Wyart,  Europhys. Lett. \textbf{30}, 387 (1995).

\bibitem{henvey}
F. S. Henyey and Y. Rabin, J. Chem. Phys. \textbf{82}, 4362 (1985).

\bibitem{rabin}
Y. Rabin, F. S. Henyey, and D. B. Creamer, J. Chem. Phys. \textbf{85}, 4696 (1986).

\bibitem{lamura}
A. Lamura, T. W. Burkhardt, and G. Gompper, Phys. Rev. E \textbf{64}, 061801 (2001).

\bibitem{gibbs}
J. W. Gibbs, 1902. \textit{Elementary principles in statistical mechanics}; Charles Scribner's Sons, New York, 1902.

\bibitem{weiner}
J. H. Weiner, \textit{Statistical mechanics of elasticity}; Dover Publication Inc., New York, 2002.

\bibitem{binder}
K. Binder, Rep. Progr. Phys. \textbf{60}, 487 (1997).

\bibitem{confser}
F. Manca, S. Giordano, P. L. Palla, F. Cleri and L. Colombo, J. Phys.: Conf. Ser. \textbf{383}, 012016 (2012).

\bibitem{cohen}
A. E. Cohen, Phys. Rev. Lett. \textbf{91}, 235506 (2003).

\bibitem{blundell}
J. R. Blundell and E. M. Terentjev, Soft Matter  \textbf{7}, 3967 (2011).

\bibitem{schwartz-math}
L. Schwartz, \textit{Mathematics for Physical Sciences}; Addison-Wesley, Reading,
MA, 1966.

\bibitem{abra}
M. Abramowitz and I. A. Stegun, \textit{Handbook of Mathematical
Functions}; Dover Publication Inc., New York, 1970.

\bibitem{grad}
I. S. Gradshteyn and I. M. Ryzhik, \textit{Table of Integrals, Series and
Products}; Academic Press, San Diego, 1965.

\bibitem{frenkel}
D. Frenkel and B. Smit, \textit{Understanding Molecular Simulation}; Academic Press, San Diego, 1996.

\bibitem{allen}
M. P. Allen and D. J. Tildesley, \textit{Computer Simulations of Liquids}; Clarendon Press, Oxford, 1987.

\bibitem{kierfeld} 
J. Kierfeld, O. Niamploy, V. Sa-yakanit, and R. Lipowsky, Eur. Phys. J. E \textbf{14}, 17 (2004).

\bibitem{rubinstein}
M. Rubinstein, R. H. Colby,  \textit{Polymer Physics}; Oxford University Press, New York, 2003.

\bibitem{kamien2}
R. D. Kamien, Rev. Mod. Phys. \textbf{74}, 953 (2002).

\bibitem{woo}
N. J. Woo, E. S. G. Shaqfeh, B.Khomami, J. Rheol. \textbf{48}, 281 (2004).



\end{thebibliography}
\end{document}